\documentclass[aps,prb,twocolumn,showkeys,amsmath,amssymb,longbibliography,floatfix,superscriptaddress]{revtex4-1}
\usepackage[T1]{fontenc}
\usepackage{graphicx}
\usepackage{dcolumn}
\usepackage{amsmath,amssymb,amsfonts,bm,dsfont,units}
\usepackage{hyperref}
\hypersetup{colorlinks=true,citecolor=blue,linkcolor=blue,urlcolor=blue,pdfstartview=FitH}
\usepackage{appendix} 
\usepackage{braket}
\usepackage{float,latexsym}
\usepackage{multirow}
\usepackage{diagbox}
\usepackage[normalem]{ulem} 
\usepackage[caption=false]{subfig} 

\newcommand{\vect}[1]{\boldsymbol{#1}}
\usepackage{array}
\usepackage{graphicx} 
\usepackage{hhline,booktabs}
\usepackage{makecell}
\usepackage{wasysym} 

\newcommand{\up}{\uparrow}
\newcommand{\down}{\downarrow}

\begin{document}

\title{Paired fermions in strong magnetic fields and daughters of even-denominator Hall plateaus}

\author{Misha Yutushui}
\affiliation{Department of Condensed Matter Physics, Weizmann Institute of Science, Rehovot 7610001, Israel}
 \author{Maria Hermanns}
 \affiliation{Department of Physics, Stockholm University, AlbaNova University Center, SE-106 91 Stockholm, Sweden}
 \author{David F. Mross}
\affiliation{Department of Condensed Matter Physics, Weizmann Institute of Science, Rehovot 7610001, Israel}

\date{\today}
	\begin{abstract}
 
 Recent quantum Hall experiments have observed `daughter states' next to several plateaus at half-integer filling factors in various platforms. These states were first proposed based on model wavefunctions for the Moore-Read state by Levin and Halperin. We show that these daughters and their parents belong to an extensive family tree that encompasses all pairing channels and permits a unified description in terms of weakly interacting composite fermions. Each daughter represents a bosonic integer quantum Hall state formed by composite-fermion pairs. The pairing of the parent dictates an additional number of filled composite-fermion Landau levels. We support our field-theoretic composite-fermion treatment by using the $K$-matrix formalism, analysis of trial wavefunctions, and a coupled-wire construction. Our analysis yields the topological orders, quantum numbers, and experimental signatures of all daughters of paired states at half-filling and `next-generation' even-denominators. Crucially, no two daughters share \textit{the same} two parents. The unique parentage implies that Hall conductance measurements alone could pinpoint the topological order of even-denominator plateaus. Additionally, we propose a numerically suitable trial wavefunction for one daughter of the SU(2)$_2$ topological order, which arises at filling factor $\nu=\frac{6}{11}$. Finally, our insights explain experimentally observed features of transitions in wide-quantum wells, such as suppression of the Jain states with the simultaneous development of half-filled and daughter states.

	\end{abstract}
	\maketitle
\section{Introduction}
\label{sec: introduction}

\begin{figure*}[t]
 \centering
\includegraphics[width=0.99\textwidth]{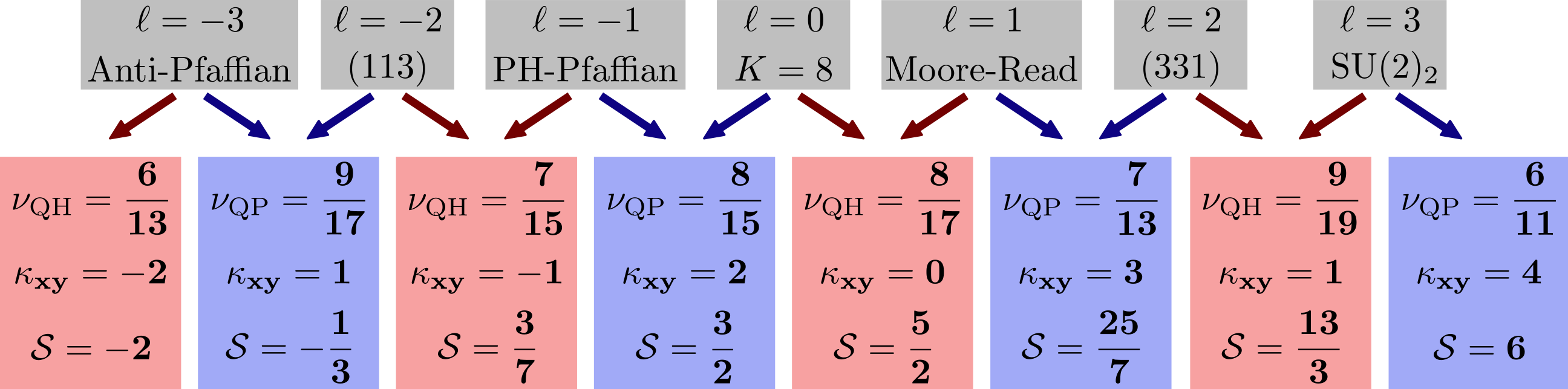}
 \caption{
 Each paired quantum Hall state at $\nu=\frac{1}{2}$ has one daughter on its hole-doped side $\nu<\frac{1}{2}$ (shaded red) and one on its particle-doped side $\nu>\frac{1}{2}$ (shaded blue). Every daughter has neighboring Abelian and non-Abelian parents. The filling factors $\nu_\text{QH}$ and $\nu_\text{QP}$ of daughter states coincide with members of the Jain sequence $\nu=\frac{n}{2n+1}$ with positive and negative $n$, respectively. However, their neutral modes differ significantly, and thermal transport or upstream noise measurements could thus distinguish between the two possibilities. The thermal Hall conductances $\kappa_{xy}$ of the daughter states differ by eight thermal conductance quanta from those of Jain states at the same filling factor. Additionally, Jain and daughter states are sharply differentiated by the shift quantum number ${\cal S}$.}
 \label{fig.daughter_states} 
 \end{figure*} 
Fractional quantum Hall states at even-denominator filling factor $\nu$ are intimately related to paired superfluids in two dimensions. This connection was already pointed out by Moore and Read shortly after he discovery of the first even-denominator plateau.~\cite{Moore_nonabelions_1991,Willett_observation_1987} It becomes explicit within Jain's composite-fermion framework.~\cite{Jain_composite-fermion_1989} Composite fermions are electrons bound to an even number $2p$ of quantized vortices. The non-zero Hall conductance implies that each vortex is associated with a quantized electric charge, which precisely cancels the one carried by electrons for $\nu=\frac{1}{2p}$. Consequently, composite fermions are neutral and insensitive to magnetic fields. As such, they can form a metallic state known as a composite-Fermi liquid (CFL).~\cite{halperin_theory_1993} Alternatively, they can form a BCS superconductor, which amounts to an incompressible quantum Hall liquid of the physical electrons.~\cite{Greiter_half_filled_1991,Wen_Non-Abelian_1991}

The composite-fermion framework maps the characterization of even-denominator quantum Hall states onto the problem of paired superfluids in two dimensions. The latter problem was analyzed in great detail in a celebrated study by Read and Green.~\cite{Read_paired_2000} In essence, fermionic superfluids with a bulk gap exhibit any integer number $\ell$ of chiral Majorana modes at their boundary. This number can be odd or even, depending on the number of internal degrees of freedom, such as valley, spin, or layer index, and we allow for both possibilities. These modes are topologically protected and stable without any symmetry. The same modes also encircle any vortex where the superconducting order parameter vanishes in the bulk. However, each vortex constitutes a zero-dimensional system where chirality ceases to be meaningful. As such, Majorana fermions encircling a vortex can gap out in pairs, and only their parity is protected.~\cite{Cheng_Tunneling_2010} 
Consequently, even $\ell$ describes Abelian states and odd $\ell$ non-Abelian states.~\cite{nayak_non-abelian_2008}

For the $\nu=\frac{5}{2}$ plateau in GaAs, numerical studies have long suggested a particular non-Abelian paired state to be realized.~\cite{Morf_transition_1998,Rezayi_incompressible_2000,Peterson_Finite_Layer_Thickness_2008,Wojs_landau_level_2010,Storni_fractional_2010,Rezayi_breaking_2011,feiguin_density_2008,Feiguin_spin_2009} Thermal transport experiments carried out over the past years have provided strong evidence in support of the so-called PH-Pfaffian phase corresponding to $\ell=-1$.~\cite{Banerjee_observation_2018,Dutta_novel_2022,Dutta_Isolated_2022} However, this non-Abelian phase has never been observed in numerics, which find the Moore-Read with $\ell =1$ or anti-Pfaffian with $\ell =-3$ instead.~\cite{Lee_particle_hole_2007,Levin_particle_hole_2007} Disorder effects may provide an avenue to reconcile numerics and experiments, but the situation remains unresolved at present.~\cite{Mross_theory_2018,Wang_topological_2018,Lian_theory_2018} 

In parallel to these developments, even-denominator plateaus have been observed in different platforms.~\cite{Ki_bilyaer_graphene_2014,Falson_Zno_2015,Kim_bilayer_graphene_2015,Li_bilayer_graphene_2017,Zibrov_Tunable_bilayer_graphene_2017,Falson_Zno_2018,Zibrov_Even_Denominator_2018,Kim_Even_Denominator_f_wave_2019,Assouline_Energy_Gap_bilayer_graphene_2024} No direct thermal measurements to identify their topological order have been performed. Nevertheless, in bilayer graphene, the non-Abelian Moore-Read or anti-Pfaffian states are suggested by satellite plateaus observed at the filling factors of the corresponding Levin-Halperin daughter states.~\cite{Levin_daughter_2009} A similar pair of plateaus suggesting the Moore-Read state was reported for wide GaAs quantum wells,~\cite{Singh_topological_2023} which exhibit a prominent plateau at $\nu=\frac{1}{2}$.

In narrow GaAs quantum wells, a single anomalous plateau consistent with anti-Pfaffian order had previously been observed.\cite{Kumar_Nonconventional_2010} More recently, a single presumed daughter of the Moore-Read has been observed in several half-filled orbitals of trilayer graphene.~\cite{chen_tunable_2023}

All these putative daughter states can be understood within the hierarchy construction pioneered by Haldane~\cite{Haldane_fqh_1983} and Halperin;~\cite{Halperin_fqh_1983} see also Ref.~\onlinecite{hansson2017quantum} for a review and several generalizations. Most pertinent for us is the work by Levin and Halperin, who investigated what phases could arise from a finite concentration of non-Abelian excitations on top of the Moore-Read state.~\cite{Levin_daughter_2009} They showed that the simplest phases accessible in this way are Abelian quantum Hall states. Crucially, the daughters of the Moore-Read and anti-Pfaffian arise at distinct electron filling factors. When present, daughter states could thus provide valuable information about the nature of the even-denominator plateau. This reasoning is somewhat complicated by the fact that the filling factors of daughter states coincide with the ubiquitous Jain states, which arise at $\nu_n=\frac{n}{2n+1}$. Jain states and daughter states are sharply distinct by their thermal Hall response, but no such measurements are available yet. Still, the anomalous strength of plateaus close to $\nu=\frac{1}{2}$ compared to lower-order Jain states supports the possibility that they are indeed daughter states.

In this article, we provide a comprehensive and unified description of the daughter states from several angles: (i) the composite fermion framework, (ii) $K$-matrix analysis, (iii) trial wavefunctions, and (iv) coupled-wire construction. Our main quantitative findings are the quantum numbers of all daughter states. Those most likely to be experimentally relevant are summarized in Fig.~\ref{fig.daughter_states}. In addition, our analysis reveals a common origin for the qualitative features observed in numerous experimental and numerical studies: (i) Suppression of Jain states in wide quantum wells and hole systems. (ii) Concomitant development of a plateau of half-filling. (iii) Appearance of daughter states flanking the plateau at half-filling.

\begin{figure}[hb]
\includegraphics[width=0.99\linewidth]{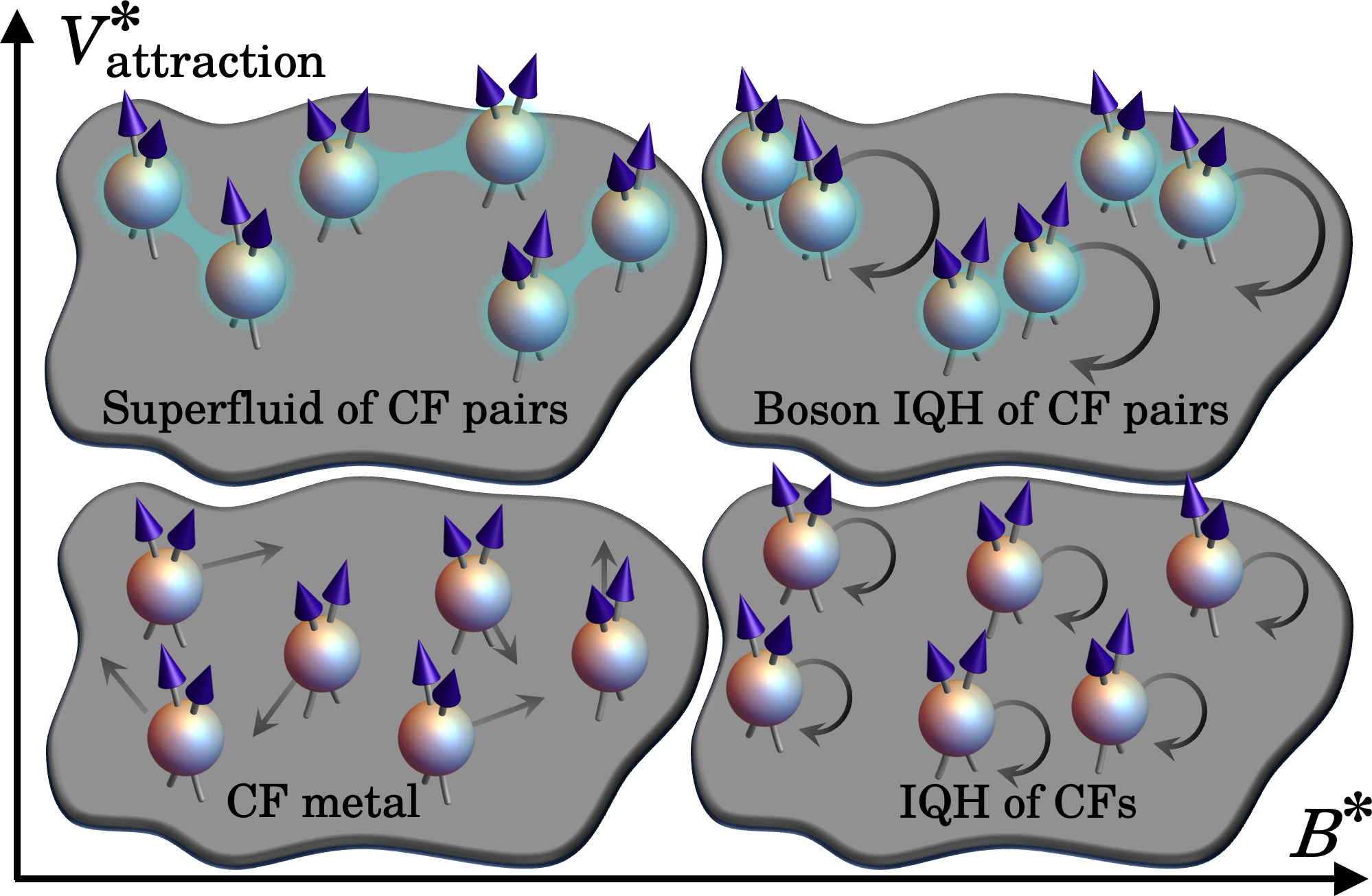}
 \caption{The phase space of the observed fractional quantum Hall state can be broadly understood in terms of the effective magnetic field $B^*$ experienced by composite fermions and their residual interaction $V^*_\text{attraction}$. When both are zero, composite fermions form a metallic CFL state. Changing the effective magnetic field results in IQH states (Shubnikov-de Haas oscillations) when $\rho_\text{CF}/B^*$ is a non-zero integer. At $B^*=0$, weak attraction can already drive an instability of the CFL into a composite-fermion superconductor. States at $B^* \neq 0$ are protected by a cyclotron gap $\omega^*_c$. When the attractions exceed this energy scale $V^*_\text{attraction}\gg\omega^*_c$, composite fermions can bind and form a quantum Hall state of bosonic pairs.}
 \label{fig.phases}
\end{figure}

Our physical picture is illustrated in Fig.~\ref{fig.phases}. The key player is the residual attraction between composite fermions. Empirically, such attraction is negligible in the narrow quantum wells with small Landau level mixing. Moreover, countless numerical studies have found excellent agreement between variational states constructed of non-interacting composite fermions and Coulomb ground states. Their excitation gap roughly follows $\Delta_n \sim \frac{E_C}{2|n|+1}$, which may be interpreted as the composite-fermion cyclotron energy $\omega^*_n$.\cite{halperin_theory_1993,Jain_composite_2007} 

The attraction between composite fermions can originate in different mechanisms that reduce the Coulomb penalty for electrons closely approaching each other. Firstly, the orbitals of the first excited Landau level are more extended in-plane and thereby suppress the short-distance repulsion compared to those of the lowest level. The lowest Landau level can benefit from this effect via Landau-Level mixing.~\cite{Zhao_mixing_2023} In experiments, Landau-level mixing is primarily controlled by the magnetic field; at a fixed filling factor, the ratio between Landau-level splitting and Coulomb energy increases with the magnetic field as $\frac{\omega_c}{E_c}\propto \sqrt{B}$ for quadratically dispersing electrons. Alternatively, hole systems feature a larger effective mass and, thus, a reduced $\omega_c$. A second mechanism is operative in wide quantum wells, where the transverse extent of the wavefunction increases with carrier density. The effective width of the quantum well thereby increases, which, in turn, reduces repulsion.~\cite{Sharma_quarter_2024}

It is generally accepted that weak residual attractions can cause gapless composite fermions in a CFL to pair up and form a BCS superconductor, resulting in the plateau at half-filling.\footnote{In reality, pairing in a CFL is more subtle due to the absence of a coherent fermion quasiparticle. Consequently, the CFL may be stable up to a finite interaction strength, unlike conventional Fermi liquids. In practice, the deviations from Fermi liquid are presumed to be logarithmic and may only complete into play for asymptotically weak interactions.~\cite{Metlitski_pairing_2015}} In parallel, the $n$th Jain state disappears once the attraction energy scale $V_\text{att}$ exceeds $\omega^*_n$. In this work we propose that the suppression of Jain states reflects a tendency of composite fermions to form pairs, analogous to the `strong pairing' of fermions when the chemical potential is within the band gap.~\cite{Read_paired_2000} However, the effective magnetic field $B^* \neq 0$ prevents such pairs from forming a superfluid. Instead, they can form a boson integer quantum Hall (IQH) state~\cite{Chen_SPT_2012,Lu_Classification_2012,Levin_Classification_2012,Ashvin_BIQH_2013,Senthil_BIQH_2013,Geraedts_realization_2013} when their density $\rho_\text{Pair}$ satisfies $\nu_\text{Pair} \equiv \frac{\rho_\text{Pair}}{B^* e_\text{Pair}}= 2$, in units where $e=h=1$ and with $e_\text{Pair}=2$. The filling factor of composite fermions with density $\rho_\text{CF} = 2 \rho_\text{Pair}$ is consequently $\nu_* \equiv \frac{\rho_\text{CF}}{B^*} = 8$. For $\nu_*=8+m$ with $m>0$, the excess composite fermions fill $m$ Landau levels. For $m<0$, there are $|m|$ filled Landau levels of composite-fermion holes. We show that the boson integer quantum Hall effect of composite fermion pairs is the incarnation of the daughter states in the composite fermion language.

The rest of this article is organized as follows: In Sec.~\ref{sec.abelian}, we analyze Abelian (even $\ell$) superfluids and their fate in a strong magnetic field. This analysis yields the filling factors, topological order, and experimental signatures of the daughter states. In Sec.~\ref{sec.anyon_condensation}, we construct $K$ matrices for all Abelian paired states and use Wen's anyon condensation approach as an alternative route to derive the daughter states. In Sec.~\ref{sec.nonabelian}, we will turn to the non-Abelian case and generalize the analysis of Levin and Halperin to other pairing channels and filling factors. 
Sec.~\ref{sec.wires} derives coupled-wire models for IQH states of Cooper pairs and for daughter states. Sec.~\ref{sec.generation} generalizes the derivation of daughter-state quantum numbers to `next-generation' even denominators, where composite fermions form a paired quantum Hall fluid instead of a superconductor. In Sec.~\ref{sec.experiment}, we discuss the observation or absence of daughter states in the most prominent experimental systems in light of the insights developed here. We conclude in Sec.~\ref{sec.conc} by discussing the implications of our results for future experimental, numerical, and analytical studies.

\section{Abelian superfluids and integer quantum Hall effect of Cooper-pairs}\label{sec.abelian}

 We begin with a two-dimensional Abelian superconductor of electrons and later attach flux.~\footnote{The analysis of this section is equally applicable to an actual superconductor.} The nature and possible phases of such superconductors were analyzed in Ref.~\onlinecite{Read_paired_2000}. They correspond to class D in the Altland-Zirnbauer classification~\cite{Zirnbauer_Riemannian_1996,Altland_Nonstandard_1997} and are identified by their thermal conductance $\kappa_{xy}=m\kappa_0$ with $\kappa_0=\frac{\pi^2 k_B^2 T}{3 h}$.~\cite{Ryu_Topological_superconductors_2010,Kitaev_Periodic_2009} The edge of such a state features $m$ complex fermions at the boundary. When subjected to a magnetic field, vortices form in this superfluid. As discussed in the introduction, they do not trap any protected zero modes at the vortex core. A finite density of such vortices can form an Abrikosov lattice without changing the topological invariant $\kappa_{xy}$ of the superfluid. In particular, the chiral modes at the boundary remain unaffected, and single-fermion excitations remain gapped throughout the bulk. The vortex density is set by the external magnetic field $\rho_\text{vortex} = B_\text{ext}$. In turn, vortices experience the charge of Cooper-pairs as magnetic flux $B_\text{vortex} = \rho_\text{pair}$. As such, the filling factors of Cooper-pair and vortices are exactly inverse
\begin{align}
\nu_\text{Pair} \equiv \frac{\rho_\text{Pair}}{B_\text{ext}}=\frac{B_\text{vortex}}{\rho_\text{vortex}} = \frac{1}{\nu_\text{vortex}}~.
\end{align}
For rational values of $\nu_\text{Pair} (\nu_\text{vortex})$, a quantum Hall phase may arise instead of the Abrikosov lattice.~\cite{Horovitz_Hall_vortex_1995}

When the external magnetic field is small $B_\text{ext}\to0$, the superfluid is recovered $\nu_\text{Pair} \to \infty$. However, for moderate magnetic fields $\rho_\text{Pair} \approx \rho_\text{vortex}$, the simplest quantum Hall state arises when $\nu_\text{Pair}=\pm2$ (or any \textit{even} integer). Here, the Cooper pairs can form a quantum Hall state without fractional excitations. This phase is equivalent to a $\nu_\text{vortex}=\frac{1}{2}$ Laughlin state of vortices.~\cite{Horovitz_Hall_vortex_1995} Crucially, it may form without closing the single-particle gap, and we can infer the nature of the resulting phase by assuming that this is the case. In this situation, the $m$ complex fermions must remain at the boundary. Moreover, the destruction of superconductivity by the coherent vortex motion restores charge conservation. Since the resulting phase is unfractionalized, they become electronic edge modes, smoothly connected to free fermion channels. The chiral modes of the boundary dictate the bulk Hall conductance to be
\begin{align}
\sigma_{xy} = (m \pm 8)\frac{e^2}{h}~.
\end{align}
The second contribution is due to the boson IQH states of Cooper pairs with $\sigma^{\nu_\text{Pair}}_{xy} = \pm 2 \frac{(e_\text{Pair})^2}{h} = \pm 8 \frac{e^2}{h}$. Its sign is determined by the direction of the magnetic field $B_\text{ext}$.

Within the $K$-matrix formalism, the resulting state is described by the $|m|+2$ dimensional block-diagonal $K$-matrix
\begin{align}\label{eq.K_daughter}
 K_{\pm} = \text{diag}(\pm\sigma_x,s,\ldots,s),\qquad 
 \vect{t} = (2,2,1,\ldots,1)~,
\end{align}
where $\sigma_x$ is a Pauli matrix, and $s=\text{sgn}(m)$ indicates the chirality of the fermions. The transport coefficients can be immediately computed from the $K$-matrix, e.g., the electric and thermal Hall conductances are $\sigma_{xy} = \vect{t} K_{\pm}^{-1} \vect{t} = m \pm 8$ and $\kappa_{xy} = m$, respectively. This phase strongly violates the Wiedemann-Franz law and is a symmetry-protected topological phase of electrons that is intrinsically interacting.~\cite{Sullivan_Interacting_2020}

In the case when the Cooper pairs and integer modes move in the same direction, i.e., when $s=\pm 1$ in the matrix $K_\pm$, the state is not topologically stable. It can undergo a phase transition to an IQH state of CF triplets, i.e., three-fermion bound states, see Sec.~\ref{sec.trion}.

\subsection{Daughter states as IQH of Cooper-pairs}\label{sec.bosoniqh}
So far, we have discussed a general possibility that could occur in any Abelian superfluid with $m$ complex fermions subjected to an external magnetic field. When this situation occurs in a superfluid formed by emergent composite fermions instead of fundamental electrons, daughter states arise. Specifically, a superfluid of composite fermions with two fluxes amounts to a quantum Hall state at half-filling. Boson IQH phases of composite-fermion pairs at $\nu^*_\text{Pair}=+2$ and $\nu^*_\text{Pair}=-2$ instead correspond to quantum Hall states at electron filling factors larger or smaller than $\nu=\frac{1}{2}$, respectively. 

From the perspective of the parent state, such fillings arise from (quasi-)hole or (quasi-)particle doping. Accordingly, we denote all daughters above half-filling by `QP' and those below by `QH.' Attaching two flux quanta transforms the composite-fermion filling factor $\nu^*=m+4\nu^*_\text{Pair} $ to the electron filling factor $\nu=\frac{\nu^*}{2 \nu^*+1}$. For the daughter states, we thus find
\begin{align}\label{eq.nu_abelian}
 \nu_\text{QH} = \frac{8 + m}{2 (8 + m)+1}~,\qquad
 \nu_\text{QP} = \frac{8 - m}{2 (8 - m)-1}~, 
\end{align}
respectively. 

The cases of QH for $m=0$ and QP for $m=1$ coincide with the first Levin-Halperin~\cite{Levin_daughter_2009} daughter states of the Moore-Read phase, obtained by condensation of quasiholes and quasiparticles, respectively.~\cite{footnote_1} The cases $m=-2$ and $m=-1$ for QH and QP naturally describe the daughter state of anti-Pfaffian obtained by particle-hole conjugation in Ref.~\onlinecite{Levin_daughter_2009}. The filling factors and relevant quantum numbers for general pairing channels are listed in Fig.~\ref{fig.daughter_states} for the first few values of $m$. All higher-descendant daughter states are obtained by larger boson IQH filling factors, as will be shown in section~\ref{sec.generation}.

\paragraph{The topological order of the daughter states}
The edge theory and the bulk quasiparticle content can be conveniently described in terms of $K$-matrix and charge vector $\vect t$.~\cite{Wen_edge_1991,Wen_shift_1992,Wen_Topological_1995} They encode the bulk filling factor as $\nu = {\vect t}^T K^{-1} \vect t$ and the edge theory via the Lagrangian density
\begin{align}
{\cal L}= \frac{1}{4\pi}\sum_{i,j}\partial_x \phi_i\left( K_{ij}\partial_t \phi_j - V_{ij}\partial_x \phi_j\right)~,
\label{eqn.klagrange}
\end{align}
where $V$ is a non-universal, positive definite velocity matrix. Excitations are encoded by integer-valued vectors $\vect l$ as $e^{i \vect l^T \vect \phi}$ and carry charge $q_{\vect l} = \vect t^T K^{-1} \vect l$.

Flux attachment is also readily performed within the $K$-matrix formalism.~\cite{Wen_shift_1992} Attaching $p$ flux quanta to the topological phase described by the K-matrix $K_{\pm}$ and charge vector $\vect{t}$ in Eq.~\eqref{eq.K_daughter}, yields the $K$-matrix 
 \begin{align}\label{eq.K_QH}
 K_\text{QH/QP}= p\,\vect{t}\vect{t}^T + K_\pm~,
\end{align}
with the same charge vector $\vect{t}$ for the daughter state topological orders.

Consequently, the daughter states occur at the same filling factor $\nu_\text{QH/QP}~=~\vect{t}^T K^{-1}_\text{QH/QP}\vect{t}$ [cf. Eq.~\eqref{eq.nu_abelian}] as the Jain states. Moreover, the minimal charge of quasiparticles is identical and is given by the denominator of their filling factor. Nevertheless, they are distinct topological orders and have various experimentally observed signatures. For example, we note that the daughters of the $K=8$ state inherit a single-electron gap on the edge. Consequently, we predict the shot noise of tunneling across vacuum into such an edge to reflect $e^*=2$. Moreover, the smallest-charge quasiparticles that exist as deconfined bulk excitations are gapped on the edge. Thus, we predict that the minimal Fano factors for tunneling across $\nu_\text{QP}=\frac{8}{15}$ and $\nu_\text{QH}=\frac{8}{17}$ correspond to $e^*=\frac{2}{15}$ and $e^*=\frac{2}{17}$. To sharply distinguish other daughter states, different experimental signatures are needed. 

\paragraph{Thermal Hall conductance}\label{sec.thermal}
Jain and daughter states could be distinguished experimentally by their thermal Hall conductance. From the $K$-matrix Eq.~\eqref{eq.K_QH}, we immediately calculate the thermal Hall conductance as the difference between the number of downstream and upstream modes,~\cite{Read_paired_2000} corresponding to positive and negative eigenvalues of $K_\text{QH/QP}$. We thus find the thermal conductance
\begin{align}
\label{eqn.kappa.daughter}
\kappa^\text{QH}_{xy}= m \kappa_0,\qquad\kappa^\text{QP}_{xy}=(m+2)\kappa_0~.
\end{align}

We contrast these values to those of Jain states at the same filling factors. At filling factor $\nu_\text{QH}$ of Eq.~\eqref{eq.nu_abelian}, a Jain state has $m + 8$ filled Landau levels of composite fermions, each contributing a single downstream mode. Attaching a positive number of fluxes does not change the number or chirality of these modes, and the thermal Hall conductance is 
\begin{align}
\kappa^\text{QH,Jain}_{xy}=(m+8)\kappa_0~.
\end{align} 
The Jain state at filling factor $\nu_\text{QP}(m)$ is particle-hole conjugate to $\nu_\text{QH}(-m-1)$. Under a particle-hole transformation $\kappa_{xy}\rightarrow 1-\kappa_{xy}$, and we obtain
\begin{align}
\kappa^\text{QP,Jain}_{xy}=(m-6)\kappa_0~.
\label{eqn.kappa.qpjain}
\end{align}
On either side of the $\nu=\frac{1}{2}$ plateau, the thermal Hall conductances of the daughter states thus deviate by eight thermal conductance quanta from those of the corresponding Jain states, i.e., $|\kappa^\text{QP/QH}_{xy}-\kappa^\text{QP/QH,Jain}_{xy}|=8\kappa_0$. This difference can be anticipated from the fact that IQH states of fermions at $\nu^*=\pm8$ have $\kappa_{xy}^*=\pm8\kappa_0$, while IQH states of bosons at $\nu^*_\text{Pair}=\pm2$ exhibit $\kappa_{xy}^*=0$. Such a significant difference in the thermal Hall conductance should be clearly visible in heat transport measurements despite uncertainties about edge-mode equilibration.
\paragraph{Upstream noise}\label{sec.noise}
An alternative signature with a smaller experimental footprint is upstream noise, which is zero for a chiral edge and non-zero in the presence of upstream modes.~\cite{Bid_Observation_neutral_2010} Conservation of energy and charge requires that any edge with $\kappa_{xy} < 1$ contains upstream modes.~\cite{Yutushui_Identifying_non_abelian_2023} For Jain states and first-generation daughters, the converse also holds, and states with $\kappa_{xy}\geq 1$ are chiral, see Sec.~\ref{sec.trion}. From Eqs.~\eqref{eqn.kappa.daughter}-\eqref{eqn.kappa.qpjain} it then follows that upstream noise distinguishes the daughter states at $\nu_\text{QH}$ with $m\leq 0$ and $\nu_\text{QP}$ with $m \geq -1$ from the corresponding Jain states. In particular, all daughters of $\nu=\frac{1}{2}$ that have been observed to date can be identified in this way.

\paragraph{Shift on a sphere} Quantum Hall states are characterized by an additional quantum number known as the shift ${\cal S}$.\cite{Wen_shift_1992}
It is related to the Hall viscosity~\cite{avron_viscosity_1995,tokatly_viscosity_2007,read_viscosity_2009} and is of vital importance for numerical studies. On a finite sphere, the magnetic flux $N_\phi$ is offset from its thermodynamic value set by $\nu$ by a factor of order unity, i.e.,
\begin{align}
{\cal S} &\equiv \nu^{-1}N_e -N_\phi~,\label{eqn.shift}
\end{align}
where $N_e$ is the number of electrons. 

The composite-fermion framework provides an intuitive avenue for obtaining the shift. Any Jain state can be viewed as $n$ filled Landau levels of composite fermions experiencing the effective flux $N_\phi^* = \pm ( N_\phi - 2(N_e-1))$. The $i$th Landau level on a sphere has $g_i=2i+|N_\phi^*|+1$ degenerate single-particle orbitals. Consequently, the total number of fermions required for filling the first $n$ Landau levels is $N^{(n)}_\text{fermions}=\sum_{i=0}^{n-1} g_i =n(n+|N_\phi^*|)$. Extracting $N_\phi$ as a function of $N_e=N^{(n)}_\text{fermions}$ yields the well-known shift of the Jain states ${\cal S} = 2\pm n$.
 
We propose that the shift of daughter states can be obtained in a similar manner. First, we note that the effective flux experienced by each pair is twice the one experienced by individual fermions, and the degeneracy of the lowest Landau level is $2|N_\phi^*|+1$. Since bosons do not have an exclusion principle, they occupy the lowest Landau level $\nu^*_\text{Pair}$ times. Thus, the number of pairs is $N_\text{pairs} = \nu^*_\text{Pair} (2|N_\phi^*|+1)$. Replacing the eight lowest Landau levels of $\nu^* = 8 + m$ by a $\nu^*_\text{Pair}=2$, the total number of electrons is 
\begin{align}
  N_e= N^{(m+8)}_\text{fermions} + (2N_\text{pairs}-N^{(8)}_\text{fermions})~.
\end{align} 
The QH or QP daughter states occur at effective flux $N_\phi^* =\pm ( N_\phi - 2(N_e-1))$, corresponding to $\nu^*=8\pm m$. Thus, the resulting shift is
\begin{align}
{\cal S}_\text{QH/QP}^{(m)}&=m+(2 \pm 8)\frac{m+2}{m\pm 8}~.
\end{align}
 For QH with $m=0$ and QP with $m=1$ and their particle hole-conjugates, QP with $m=-1$ and QH $m=-2$, our expressions reproduce results derived by Levin-Halperin.~\cite{Levin_daughter_2009} Unfortunately, almost all daughters of the most important paired states have fractional shifts (cf.~Fig.~\ref{fig.daughter_states}) and thus can not be represented by a numerically efficient parton wavefunction. Notable exceptions are $\nu_\text{QH}=\frac{6}{13}$ studied before,~\cite{Balram_parton_6_13_2018} and $\nu_\text{QP}=\frac{6}{11}$, for which we propose a wavefunction in Sec.~\ref{sec.conc}. By contrast, the daughters of higher pairing channels have predominantly integer shifts.

\subsection{Integer Quantum Hall State of composite fermion triplets}\label{sec.trion}
 
\begin{figure}[ht]
 \centering 
 \includegraphics[width=0.90\linewidth]{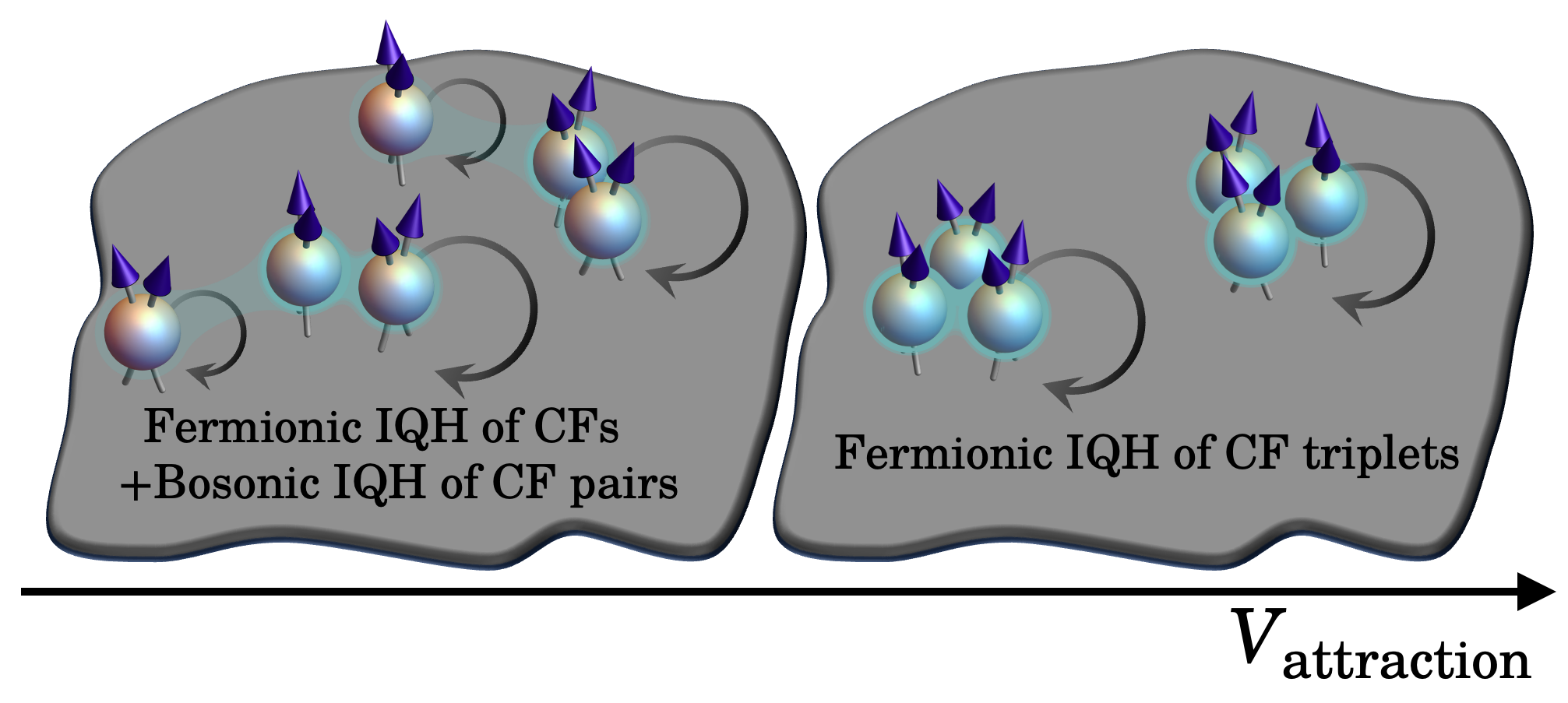}
 \caption{A state comprised of a fermionic IQH liquid layered on top of a bosonic IQH liquid of fermion pairs with the same chirality is topologically unstable. It reduces to an IQH state of composite fermion triplets, bound states of three composite fermions.}
 \label{fig.trio}
\end{figure} 

In Sec.~\ref{sec.bosoniqh}, we developed a description of daughter states as IQH phases of composite-fermion Cooper pairs. Here, we show that the topological order is unstable when there are additional fermionic IQH quantum Hall states with the same chirality. Each pair can bind with an additional unpaired fermion to form fermionic triplets with charge three. Triplets, in turn, can form an IQH state when $\nu^*_\text{triplet} = \nu^* /9$ is an integer. 

The simplest case when this situation arises is for the electron-filling factors $\nu_\text{QP}=\frac{9}{19}$ and $\nu_\text{QH}=\frac{9}{17}$. The latter filling factor has been observed experimentally in bilayer graphene Ref.~\onlinecite{Huang_Valley_bilayer_graphene_2022}. Both states can be described by $\nu^*_\text{pair}=\pm 2$ with $m=\pm 1$ Eq.~\eqref{eq.K_daughter} with two fluxes attached. The composite fermion $K$-matrix 
\begin{align}
 K^*=\begin{pmatrix}
 0 & 1 & 0 \\
 1 & 0 & 0\\
 0 & 0 & 1\\
 \end{pmatrix}~, 
\end{align}
with charge vector $\vect{t}^T=(2,\;2,\;1)$, is topologically unstable according to Haldane criterion.~\cite{Haldane_Stability_1995} To find the reduced $K$ matrix, we perform an $\text{SL}(3,\mathbb{Z})$ transformation $K'=W^T K^* W$ and $\vect{t}' = W^T\vect{t}$ with 
\begin{align}
 W=\left(
\begin{array}{ccc}
 1 & 1 & 1 \\
 0 & -2 & -1 \\
 1 & 2 & 1 \\
\end{array}
\right)~,\quad K'=\begin{pmatrix}
 1 & 0 & 0 \\
 0& 0 & -1\\
 0 & -1 & -1\\
 \end{pmatrix}~,
\end{align}
and $\vect t^{\prime T} = (3,0,1)$. In this basis, the null-vector $\vect l  ^T = (0,0,1)$ satisfies Haldane conditions $\vect{l}^TK'^{-1} \vect{t}'=\vect{l}^TK'^{-1} \vect{l}=0$. 
The last two entries of $K'$ describe a topologically trivial pair of modes. Dropping it reduces the $K$-matrix to a single element $K_\text{red}=1$ and $\vect{t}_\text{red} = 3$, which we interpret as the IQH effect of the charge-three objects, which we call triplets. 

The electron topological order can be obtained by alternative routes. Firstly, by performing flux attachment and finding a null-
vector to reduce the $K$-matrix. Secondly, by attaching two fluxes to the reduced $K^*=\pm K_\text{red}$. Either way, we arrive at 
\begin{align}
  K^\text{triplet}_\text{QP} = 19~,\qquad K^\text{triplet}_\text{QH} = 17~,
\end{align}
with $\vect{t}=3$, which are the Laughlin states of $3e$ particles. 

This transition has immediate experimentally observable consequences,~\cite{Kao_Binding_1999,Overbosch_binining_2008,Spanslatt_binding_2023,Yutushui_Identifying_non_abelian_2023} the most prominent of which is the absence of upstream noise, see Sec.~\ref{sec.noise} and Appendix~\ref{app.noise}, and tunneling shot noise. Similarly to the $K=8$ state and its daughters, single electrons are gapped on the edge and cannot tunnel. In the IQH state of triplets, electron pairs are also gapped. Consequently, the shot noise of tunneling through vacuum reflects $e^*=3$, similar to the cases of Ref.~\onlinecite{Kao_Binding_1999,Spanslatt_binding_2023}. 

\section{Anyon condensation approach}\label{sec.anyon_condensation}

\begin{figure}[hb]
 \centering 
 \includegraphics[width=0.95\linewidth]{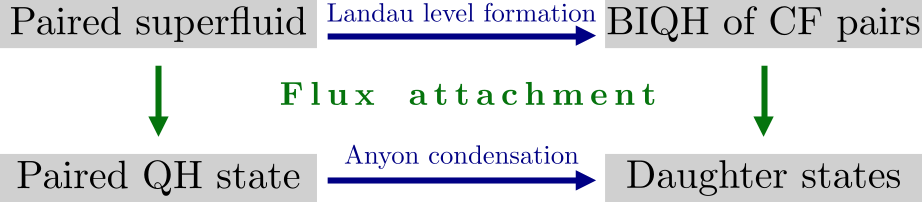}
 \caption{
 Two alternative approaches to deriving daughter states are illustrated by a commutative diagram. In Sec.~\ref{sec.abelian}, we start with the superfluid and include a magnetic field (horizontal arrow) to access the bosonic IQH effect of pairs. Attaching fluxes (vertical arrow) then yields the daughter states. By contrast, in Secs.~\ref{sec.anyon_condensation} and \ref{sec.nonabelian}, we first generate paired quantum Hall states by attaching fluxes to a superfluid. We then use anyon condensation~\cite{Wen_Topological_1995,Levin_daughter_2009} to obtain the daughter states.}
 \label{fig.anyon_cond}
\end{figure} 

In Sec.~\ref{sec.abelian}, we first constructed the daughter states of paired superfluids and then attached fluxes to obtain the corresponding quantum Hall state. An equivalent route is the anyon condensation approach of Wen.~\cite{Wen_Topological_1995} Following this procedure, we start with the $K$ matrix of any Abelian quantum Hall state and then `condense' its fundamental anyons, i.e., place them into a bosonic Laughlin state. 

 To find the $K$-matrix for paired quantum Hall states with general $\ell=2m$, we begin with a chiral superfluid described by $K_\text{SF}=\mathbb{I}-\sigma_x$, $\vect{t}^T_\text{SF}=(1,1)$, where $\mathbb{I}$ is the two-dimensional identity matrix. It can be viewed as a $11\overline{1}$ bilayer state represented by 
\begin{align}
\Psi_{11\overline{1}} = \frac{\prod_{i<j}(z_i-z_j) \prod_{i<j}(w_i-w_j)}{\prod_{i,j}(z_i-w_j)}=\det{\frac{1}{z_i-w_j}}~.
\end{align}
This wavefunction describes an Abelian weak-pairing superconductor~\cite{Read_paired_2000} with two Majorana fermions at the edge. For general $m$, we add $m-1$ filled Landau levels, where negative numbers mean Landau levels of holes. These additional levels are encoded by the $|m-1|$ dimensional matrix $K^{m-1}_\text{IQH} =\text{sgn}(m-1)\text{diag}(1,\ldots,1)$ and vector $\vect{t}_\text{IQH}=(1,\ldots,1)$. Upon attaching $p$ flux quanta, we thus obtain ${K}_m = p\,\vect{t}\vect{t}^T + K^*_m$ for the paired quantum Hall state with
\begin{align}\label{eq.K_ell_0}
K^*_m = \text{diag}(K_\text{SF}, K^{m-1}_\text{IQH}),\quad \vect{t}_{m}^T=(1,\ldots,1)~.
\end{align}
In general, this $K$ matrix is topologically unstable and admits reduction to a $K$ matrix with total dimensions $\text{dim}(K)=|m|+1$. This reduction does not affect the quasiparticle content, which matches the one given in Ref.~\onlinecite{Ken_sixteenfold_2019}.

The $K$-matrix in Eq.~\eqref{eq.K_ell_0} accurately represents the topological order for any $m$. However, the $K$-matrix in Eq.~\eqref{eq.K_ell_0} is not topologically stable for negative $m$; the edge state it describes via Eq.~\eqref{eqn.klagrange} contains modes that can localize;~\cite{Haldane_Stability_1995} see also Appendix~\ref{app.K_SF}. It is often desirable to avoid such spurious, non-topological modes. A simple way to obtain the reduced $K$-matrix, in this case, is to start with a superfluid of opposite (negative) chirality and add $m+1 \leq 0$ filled Landau levels. For $m<0$, the superfluid and integer levels have equal chirality, and there are no spurious modes. A minimal $K$ matrix valid for any $|m|\geq1$ is thus given by ${K}_m = p\,\vect{t}\vect{t}^T + K_m^*$ with
\begin{align}\label{eq.K_ell}
K^*_m = \text{sgn}(m)\; \text{diag}(K_\text{SF}, K^{|m|-1}_\text{IQH})~,
\end{align}
where $\vect t_m = (1,\ldots,1)$. In particular, Eq.~\eqref{eq.K_ell} with $p=2$ yields the familiar 331 and 113 topological orders for $m=1$ and $m=-1$, respectively. We conclude that $K_{m}$ describes any Abelian quantum Hall state at $\nu=\frac{1}{2p}$ with arbitrary even pairing $\ell=2m \neq 0$. The $m=0$ case corresponds to the $K=0$ topological order; see Appendix~\ref{app.K_SF}.

\paragraph{Daughter states from anyon condensation} The daughter states can be obtained from $K_m$ following the standard hierarchy construction.\cite{Wen_Topological_1995} The condensation of a charge-$\frac{1}{4p}$ quasiparticle represented by integer-valued vector $\vect{l}=(1,0,\ldots)$ results in 
\begin{align}\label{eq.K_daughter_AC}
K_\text{QH/QP} = 
\begin{pmatrix}
K_m & \vect{l}\\
 \vect{l}^T & \mp 2\\
 \end{pmatrix}~, \qquad \vect{t}_\text{QH/QP} = (1,\ldots,1,0)~.
\end{align}
 The resulting topological order matches the one derived in Sec.~\ref{sec.abelian}; an $SL(|m|+2,\mathbb{Z})$ transformation maps the $K$-matrices of Eq.~\eqref{eq.K_QH} and Eq.~\eqref{eq.K_daughter_AC} onto one another. The main difference between the two derivations lies in the order in which we condense anyons and attach fluxes; see Fig.~\ref{fig.anyon_cond}.

\section{Non-Abelian states}\label{sec.nonabelian}

The line of reasoning we pursued in the previous section faces several immediate challenges when confronted by non-Abelian states. Most importantly, vortices carry fermionic zero modes and thereby spoil the single-particle gap. Hybridization of nearby zero modes leads to bands of low-energy Majorana fermions.~\cite{Grosfeld_Electronic_2006,Kraus_Majorana_2011,Ludwig_Two_dimensional_2011,Biswas_vortexlattice_2013,Schirmer_Phase_diagram_2022} Generically, these bands are gapped and exhibit an odd Chern number ${\cal C}$. The corresponding edge modes combine with those of the pristine superfluid. The total number of edge Majoranas is even, indicative of an Abelian state. Unfortunately, there is no general principle determining the Chern number of the Majorana band. In particular, the possible values ${\cal C}=\pm 1$ appear equally likely and will be selected by microscopic details. The resulting Abelian phase is thus characterized by either $\ell+1$ or $\ell-1$.

When vortices form a quantum Hall state instead of a lattice, there is a preferred chirality that breaks the tie between the most probable values ${\cal C}=\pm 1$. A unique answer emerges in the limit of a single Landau level. This problem was studied by Levin and Halperin on the basis of trial wavefunctions for $e/4$ quasiparticles and quasiholes on top of the Moore-Read state.~\cite{Levin_daughter_2009} They found that the holomorphic structure naturally splits the difference. The daughters of $\ell =1$ coincide with those of $\ell=0$ or $\ell=2$ for opposite signs of the effective magnetic field. We now generalize their analysis to any odd $\ell$.

\subsection{Quasihole daughter state}
Levin-Halperin in Ref.~\onlinecite{Levin_daughter_2009} constructed a hierarchy of states on top of the $\nu=\frac{1}{2}$ Moore-Read state. Non-Abelian $e/4$ quasiholes at a finite density can form a quantum Hall state of their own. This state is constructed to respect the braiding properties of the $e/4$ quasiparticles. Crucially, the braiding depends on the pairing channel of the parent state. Consequently, the valid quasiparticle wavefunctions occur at different filling factors set by the pairing channels. 

We assume that the number of Majoranas $\ell=2m+1$ is odd. Then, there are $\ell$ electron operators $\psi_{e,l}(z)=\gamma_{l}(z)e^{i\sqrt{2}\varphi(z)}$, with $\varphi(z)$ responsible for the charge and normalized such that $\langle\varphi(z)\varphi(z')\rangle=-\log(z-z')$. It is convenient to use a different basis where $2m$ Majoranas are combined to form complex fermions $e^{\pm i\phi_k}$ with $k=1,\ldots,m$, where $\phi_k$ has the same normalization as $\varphi$. Now, the electron operators are
\begin{align}\label{eq.electron_ops}
 \psi_{e,0}(z) =\gamma e^{i\sqrt{2}\varphi},\quad 
 \psi_{e,\pm k}(z) =e^{i\sqrt{2}\varphi \pm i\phi_k}.
\end{align}
The valid $e/4$ quasiparticle operators are mutually local with all electrons and can be written as~\cite{Ken_sixteenfold_2019}
\begin{align}\label{eq.chi}
 \chi_{\vect{s}}(w) = \sigma(w) e^{i\varphi(w)/\sqrt{8}} \prod^{m}_{k=1}e^{i s_k \phi_k(w)/2}~,
\end{align}
where $s_k=\pm 1$ parameterize the type of quasiparticle. For $m=0$, there is a single $\sigma$ quasiparticle type and no ambiguity in condensing $e/4$ quasiparticle. In general, for positive $m$, there are $2^{m}$ different quasiparticle types corresponding to different sets of $s_k=\pm1$. The daughter states are obtained by condensing $\chi_{\vect{s}}(w)$ with {\it the same} $\vect{s}$ into quantum Hall states.~\footnote{The situation where quasiparticles of different types are condensed results in different anyonic filling factors and will be considered elsewhere.} Comparing the previously studied $m=0$ case Ref.~\onlinecite{Levin_daughter_2009} to the general $m$ case, we propose that a suitable quasihole wavefunction can be written as a conformal field theory (CFT) correlator of 
\begin{align}\label{eq.qh_op}
 \bar{\xi}_{\vect{s}}(\bar{w}) = \bar{\sigma}(w) e^{i\,\sqrt{2n+\frac{1}{8}}\bar{\varphi}(\bar{w})}\prod^{m}_{k=1}e^{i s_k \bar{\phi}_k(\bar{w})/2}~.
\end{align}
Analogously to Eqs.~(5),(6) of Ref.~\onlinecite{Levin_daughter_2009}, the complex conjugate of the correlation function of Eq.~\eqref{eq.qh_op} has the same transformation properties under particle exchange as the electron wavefunction with quasihole insertion Eq.~\eqref{eq.chi}. We use $\sum_{k=1}^{m}s_k^as^b_k=\sum_{k=1}^{m}s_k^2=m$ to evaluate the correlator 
\begin{align} 
\left\langle
 \prod_{a}\bar{\xi}_{\vect{s}}(\bar{w}_a)
 \right\rangle_\alpha = 
 \langle\bar{\sigma}(\bar{w}_1)\ldots\rangle_\alpha \prod_{a<b}\bar{w}_{ab}^{2n+\frac{1}{8} + \frac{m}{4}}~.
\end{align}
From this expression, we deduce the anyon filling factor 
\begin{align}\label{eq.nu_qha}
 \nu_\text{QH}^\text{anyon} = \frac{1}{2n+\frac{\ell}{8}} \qquad \text{(Anyon filling)}.
\end{align}
The electron filling factor is obtained by noting that charge density is reduced $\rho = \rho_0 - q\rho_\text{QH}$ due to the presence of quasiholes, where $\rho_0=\frac{1}{2}B$ is the unperturbed electron density, $q=\frac{1}{4}$ is the charge of quasiholes in units of electron charge, and $\rho_\text{QH} = \nu^\text{anyon}_\text{QH} B^*$ is the number density of quasiholes. Quasiholes feel an effective magnetic field $B^*=qB$. Thus, the electron filling factor $\nu_\text{QH}=\frac{\rho}{B}$ of daughter states obtained by quasihole condensation is
\begin{align}\label{eq.nu_qh}
 \nu_\text{QH}= \frac{1}{2} - \frac{1}{16}\nu_\text{QH}^\text{anyon} \qquad \text{(Electron filling)}.
\end{align} 
This result coincides with the filling factor $\nu_\text{QH}$ in Eq.~\eqref{eq.nu_abelian}, found using the composite fermion picture for Abelian states, see Fig.~\ref{fig.daughter_states}. The same analysis also extends to negative $m$. The only difference is the chirality of the neutral sector. Thus, we take $|m-1|$ holomorphic bosons $\phi_k(\omega)$ and the holomorphic part of the Ising CFT, compared to the anti-holomorphic CFTs for positive $m$. The resulting filling factors Eqs.~\eqref{eq.nu_qha} and \eqref{eq.nu_qh} are valid for both positive and negative $m$.

\subsection{Quasiparticle daughter state}
In analogy with the Levin-Halperin analysis, we chose the wavefunction of quasiparticles to be the CFT correlator of 
\begin{align}
 \rho_{\vect{s}}(w) = \sigma'(w) e^{i\,\sqrt{2n-\frac{3}{8}}\varphi'(w)} \prod_{k=1}^{m} e^{is_k\bar{\phi}_k(\bar{w})/2}~.
\end{align}
Notice that the last term is the same as in Eq.~\eqref{eq.qh_op} since the `reversion' of the monodromy is taken care of by the exponent of the first term.~\cite{Levin_daughter_2009} 
Thus, the quasiparticle wavefunction is constructed of the correlation functions 
\begin{align}
\langle\prod_{a}\rho_{\vect s}(w_a)\rangle_\alpha = 
\langle\sigma'(w_1)\ldots\rangle_\alpha \prod_{a<b} w_{ab}^{2n-\frac{3}{8}} \bar{w}^{\frac{m}{4}}_{ab}~.
\end{align}
To obtain a lowest Landau level wavefunction, we replace $\bar{w}^{\frac{m}{4}}_{ab}~\to~w^{-\frac{m}{4}}_{ab}$, expecting that $\prod_{a<b}|w_{ab}|^\frac{m}{4}$ do not significantly affect the wavefunction after projection. The anyonic filling factor is then given by
\begin{align}\label{eq.nu_qpa}
\nu^\text{anyon}_\text{QP}=\frac{1}{2n - \frac{\ell+2}{8}} \qquad \text{(Anyon filling)}.
\end{align}
 The corresponding electron filling factor of the daughter state obtained by condensation of quasiparticles is 
\begin{align}\label{eq.nu_qp}
 \nu_\text{QP} = \frac{1}{2} + \frac{1}{16}\nu_\text{QP}^\text{anyon} \qquad \text{(Electron filling)}.
\end{align}
This filling factor reproduces the filling obtained from the composite fermion perspective in Eq.~\eqref{eq.nu_abelian} for $\ell=2m-1$, see Fig.~\ref{fig.daughter_states}, and applies to positive and negative $m$. Finally, in Appendix~\ref{app.non-abelian}, we study daughter states that arise from general $2p$-flux composite superfluid at $\nu=\frac{1}{2p}$.

\section{Wire construction}\label{sec.wires}
The coupled wire formalism allows us to express superconductors, integer, and fractional quantum Hall states in a unified framework. We consider an array of identical wires aligned in the $\hat x$ direction and enumerated by integers $y$. Each wire hosts two flavors $\sigma=\uparrow,\downarrow$ of electrons, which we expand according to
$c_\sigma \sim e^{i k_{F,\sigma }x} \psi^R_{\sigma}+e^{-i k_{F,\sigma } x} \psi^R_{\sigma}$ with $k_{F,\sigma }$ the Fermi momenta. The linearized Hamiltonian of this array is ${\cal H}_\text{kin} = \sum_{y} \int_x H_\text{kin} $, with
\begin{align}\label{eq.H_nonint}
H_\text{kin} = v_F\sum_\sigma\left([\psi^R_{\sigma,y }]^\dagger i \partial_x \psi^R_{\sigma,y}-[\psi^L_{\sigma,y}]^\dagger i \partial_x \psi^L_{\sigma,y}\right)~.
\end{align}
Supplementing $H_0$ with suitable inter-wire terms can realize a wide variety of different phases. They are typically analyzed within the framework of Abelian bosonization. For our purposes, it is sufficient and more transparent to remain with the fermionic formulation. A treatment within bosonization is presented in Appendix~\ref{app.wire}.

\paragraph{Abelian superconductors} To realize various types of superconductors, we include
\begin{align}
\label{eqn.hpair}
H_\text{pair}= g \hat\Delta_{y}^\dagger \hat\Delta_{y+1} e^{i 2 b x}+\text{H.c.}~,
\end{align}
where $\hat\Delta_{y}= \psi^R_{\uparrow,y}\psi^L_{\downarrow,y}$ is a Cooper-pair operator, and $b$ is the average magnetic field in the gauge $A_y = - b x$. When $b=0$, Cooper pairs can hop across wires coherently, and a superfluid with $\langle \hat\Delta_y \rangle \neq 0$ arises. Still, to include vortices and restore the charge conservation, we must refrain from making a mean-field approximation.

The model $H_0+H_\text{pair}$ does not involve the modes $\psi^R_{\downarrow,y},\psi^L_{\uparrow,y}$, which, therefore, remain gapless. We couple these to $\hat \Delta$ via Josephson-type coupling
\begin{align}
\label{eqn.hm}
H_m=g_\text{m}\hat\Delta_{y}^\dagger \psi^R_{\downarrow,y}\psi^L_{\uparrow,y+m} e^{i(2k_{F,\uparrow}-2k_{F,\downarrow}+ m b) x }+\text{H.c.}
\end{align}
The resulting phase is gapped up to a Goldstone mode and hosts $m$ chiral edge modes. An easy way to see this is to replace $\hat 
\Delta$ with its expectation value. We also note that flux attachment to $H_\text{kin} + H_\text{pair} + H_m$ yields wire models whose ground states match those of the $K=8$ state $(m=0)$ and $331$ state $(m=1)$ derived by Teo and Kane in Ref.~\onlinecite{Teo_LL_to_non_Abelian_2014}; see also Appendix~\ref{app.wire}.

\paragraph{Integer quantum Hall state of Cooper-pairs} To instead access the boson IQH effect of Cooper pairs, we require an expression for the superconducting $\frac{hc}{2e}$ vortex. Following the logic of Refs.~\onlinecite{Mross_explicit_duality_2016,Mrossduality2017,fuji_wires_2019,leviatan_wires_2020}, we define the hopping of such a vortex across two wires via the $2k_F$ process
\begin{align}
\hat V^\dagger_{y-\frac{1}{2}}\hat V_{y+\frac{1}{2}}&=[\psi^L_{\downarrow,y}]^\dagger \psi^R_{\downarrow,y} e^{2 i k_{F,\downarrow} x}~.
\end{align}
In the boson IQH phase, the Cooper pair and vortex form a composite that condenses into a superfluid. To capture this phase, we replace $H_\text{pair}$ by
\begin{align}
H_\text{BIQH} = 
(\hat\Delta_{y-1}\hat V^\pm_{y-\frac{1}{2}})^\dagger (\hat V_{y+\frac{1}{2}}^\pm\hat\Delta_{y+1} )e^{\pm 2 i k_{F,\downarrow} x}e^{i 4 b x}
+\text{H.c.} \label{eqn.bosonvortex}
\end{align}
Crucially, this interaction does not disrupt $H_m$ and we can use $H_\text{BIQH}$ independent of $m$. 

To form a gapped state, the phase factors in $H_m$ and $H_\text{BIQH}$ must vanish. This requirement translates into a condition on the densities $\rho_\sigma = 2 k_{F,\sigma}$ in relation to the magnetic field. We find the conditions $\rho_\uparrow - \rho_\downarrow = m b$ and
\begin{align}
\nu \equiv \frac{\rho_\uparrow + \rho_\downarrow}{B} &=m \mp 8~.
\end{align}
The filling is precisely the one determined in Sec.~\ref{sec.abelian}. We emphasize that for all $m$, we condense \textit{the same} vortex-Cooper pair composite via $H_\text{BIQH}$. The filling factor is automatically determined by $H_m$
for all $m$, which encodes the number of chiral edge modes.

\section{Higher generations}\label{sec.generation}

In previous sections, we have derived the first-level daughter states of paired quantum Hall plateaus at half-filling. Here, we derive the higher descendant daughter states. Additionally, we generalize the analysis to other even-denominator states, which were recently observed.~\cite{Wang_even_3_4_2022,Wang_Next_generation_2023}

\subsection{Higher-level daughter states}
We now allow for general even filling factors of Cooper-pairs, i.e., $\nu_\text{Pair}=\pm 2n_\text{Pair}$. A bosonic quantum Hall state at this filling factor corresponds to a Laughlin state of vortices at filling factor $\nu_\text{vortex}=\frac{1}{2 n_\text{Pair}}$.
Similarly, to the IQH of fermions, the $K$-matrix can be constructed by stacking $n_\text{Pair}$ copies of $\sigma_x$ corresponding to $K$-matrix of $n_\text{Pair}=1$. Unlike the fermion case, the resulting state 
\begin{equation}\label{eq.K_daug_higher}
\begin{split}
 K_{n_\text{Pair}}= \text{diag}(\sigma_x,\ldots,\sigma_x),\qquad
 \vect{t}= (2,\ldots,2)~
\end{split}
\end{equation}
is topologically unstable. The $K$-matrix $K_{n_\text{Pair}}$ reduces to the two-dimensional
\begin{align}\label{eq.K_daug_higher_red}
     K_{n_\text{Pair}} =\begin{pmatrix}
         0 & 1\\
         1 & 2(1-n_\text{Pair})
     \end{pmatrix},\qquad \vect{t} &= (2,2)~,
\end{align}
as we verified inductively in Appendix~\ref{app.K_BIQH}; see also Refs.~\onlinecite{Lu_Classification_2012,Senthil_BIQH_2013,Ashvin_BIQH_2013}. Taking into account integer modes corresponding to a pairing channel, a straight forwards generalization of Eq.~\eqref{eq.K_daughter} reads
\begin{align}\label{eq.K_daughter}
 \begin{split}
  K^*_{\pm} = \text{diag}(\pm K_{n_\text{Pair}},s,\ldots,s)~,\\ 
  \vect{t}^* = (2,2,1,\ldots,1)~,
\end{split}
\end{align}
for composite fermion filling factor $\nu^*=m\pm 8n_\text{Pair}$. To obtain the electron topological order of the higher-level daughter states, we perform flux attachment according to Eq.~\eqref{eq.K_QH}. The resulting filling factors are given by $\nu_\text{QH/QP}=~\frac{ m\pm 8n_\text{Pair}}{2( m\pm 8n_\text{Pair})+ 1}$. This analysis reproduces the Levin-Halperin results for the Moore-Read pairing by taking QH with $m=0$ and QP with $m=1$. Similarly, the anti-Pfaffian daughters are obtained from QP with $m=-1$ and QH with $m=-2$.

\subsection{Higher-level daughter states}
We now allow for general even filling factors of Cooper-pairs, i.e., $\nu_\text{Pair}=\pm 2n_\text{Pair}$. A bosonic quantum Hall state at this filling factor corresponds to a Laughlin state of vortices at filling factor $\nu_\text{vortex}=\frac{1}{2 n_\text{Pair}}$. The resulting state is described by the $(|m|+2n_\text{Pair})$-dimensional $K$-matrix and charge vector
\begin{equation}
\begin{split}
 K^*_\pm &= \text{diag}(\pm\sigma_x,\ldots,\pm\sigma_x,s,\ldots,s),\\ 
 \vect{t}^* &= (2,\ldots,2,1,\ldots,1)~,
\end{split}
\end{equation}
in a straightforward generalization of Eq.~\eqref{eq.K_daughter}.

To obtain the electron filling factors of the higher-level daughter states, we perform flux attachment according to Eq.~\eqref{eq.K_QH}. The resulting filling factors are given by $\nu_\text{QH/QP}=~\frac{ m\pm 8n_\text{Pair}}{2( m\pm 8n_\text{Pair})+ 1}$. This analysis reproduces the Levin-Halperin result for the Moore-Read pairing by taking QH with $m=0$ and QP with $m=1$. Similarly, the anti-Pfaffian daughters are obtained from QP with $m=-1$ and QH with $m=-2$.

\subsection{Daughters of next-generation even-denominators states}\label{sec.next_gen}

Recent experiments in hole-doped GaAs observed even-denominator states at $\nu=\frac{3}{4},\frac{3}{8}$ and several other filling factors with even denominators larger than two.~\cite{Wang_even_3_4_2022,Wang_Next_generation_2023} Plateaus at these `next-generation' (NG) filling factors cannot be obtained by attaching flux to electrons and pairing the resulting composite fermions. Instead, flux attachment necessarily results in fractional composite-fermion filling factors. Attaching two flux quanta to electrons at filling factors $\nu=\frac{3}{8},\frac{3}{4}$ yield $\nu^*=\pm\frac{3}{2}$, respectively. More generally, the electron filling factors
\begin{align}\label{eq.nu_flux}
 \nu^\text{NG}(n,p) = \frac{n+\frac{1}{2}}{p\left(n+\frac{1}{2}\right) \pm 1}~.
\end{align}
yield composite fermions at half-integer filling factors $\nu^*=\pm\left(n+\frac{1}{2}\right)$ upon attaching an even number $p$ of flux quanta. 

A plateau at filling factor $\nu^\text{NG}(n,p)$ arises when the composite fermions in the $(n+1)$th Landau level form a paired quantum Hall state with filling factor $\frac{1}{2}$ and pairing $\ell$. Its daughters occur when the composite fermions in the partially filled level are at filling $\nu_\text{QH/QP}$ given by Eq.~\eqref{eq.nu_abelian}. The electron filling factors of the daughters are consequently given by
\begin{align}
\nu^\text{NG}_\text{QH/QP} =
 \frac{n+\nu_\text{QH/QP}}{p(n+\nu_\text{QH/QP}) \pm 1}~.
\end{align}
This formula reduces to the half-filled case for $n=0$ and $p=0$ with the positive choice of sign. Any non-zero, even $p$ instead describes the daughters of the paired state at $\nu=\frac{1}{2p+2}$. In Appendix~\ref{app.higher}, we provide several examples for which we obtain the same filling factors by explicit calculation and determine the $K$ matrices.

\section{Experimental implications}\label{sec.experiment}
We now comment on the most prominent experimentally observed even-denominator plateaus from the perspective developed here. In particular, we address the relative strength or absence of daughter states in some cases, their competition with Jain states, and disorder effects. Finally, we discuss the expected daughters of the leading `next-generation' even denominator states in quarter-filled Landau levels. 

\subsection{Jain--Daughter state transitions in wide quantum wells}
Recently, transitions between Jain states and daughter states that occur upon tuning the electron density in wide quantum wells have been reported in Ref.~\onlinecite{Singh_topological_2023}. The authors observed that the plateaus at the filling factors of the Moore-Read daughters, $\nu_\text{QH} = \frac{8}{17}$ and $\nu_\text{QP}=\frac{7}{13}$, are initially suppressed and then enhanced with increasing density or in-plane magnetic field. They interpret this finding as Jain states at low density that undergoes topological phase transitions to daughter states. 

This interpretation is supported by our findings that daughter states result from the formation of composite fermion pairs that realize a quantum Hall state. Specifically, the short-range repulsion between electrons in wide quantum wells becomes weaker with increasing density or in-plane field.~\cite{Park_width_1998,Hasdemir_tilted_magnetic_2015,Jungwirth_Self_1993} As a consequence, the residual attraction $V^*_\text{attraction}$ between composite fermions becomes stronger. In general, this interaction strength also depends on the filling factor. However, a simplified model with a single filling-factor-independent composite-fermion interaction suffices to qualitatively account for the observations.

At half-filling, weak attraction drives a superconducting instability of the gapless composite Fermi sea.~\cite{Metlitski_pairing_2015} Further increasing $V^*_\text{attraction}$ leads to a better-developed plateau at half-filling but does not induce any transitions. By contrast, Jain states are protected by a finite energy gap, the composite-fermion cyclotron energy $\omega_{n}^*\approx \frac{E_c}{2n+1}$. Once $V^*_\text{attraction}$ exceeds the energy gap of the Jain state at the same filling factor $ \omega_{n_0} \lesssim V^*_\text{attraction}$, the phase transition to the daughter state takes place. Further increase of the attraction strength washes out of lower order Jain states when their cyclotron gap is overcome by attraction $ V^*_\text{attraction}\gtrsim\omega_{n<n_0} $.

\subsection{Single daughter states}
In some experiments, only a single daughter is observed. Ref.~\onlinecite{Kumar_Nonconventional_2010} found a plateau consistent with the anti-Pfaffian daughter at $\nu_\text{QH}=\frac{6}{13}$ in the second Landau level of GaAs. Similarly, Ref.~\onlinecite{chen_tunable_2023} found a state at the filling factor of the Moore-Read daughter $\nu_\text{QP} = \frac{7}{13}$ flanking the half-filled plateau of multiple Landau levels in trilayer graphene. In both cases, the absence of lower-order Jain states supports the interpretation of these plateaus as daughter states. Their appearance thus narrows down the potential pairing channels of the parent states.

However, a single daughter cannot differentiate between the topological orders of her two parents. As such, observing the $\nu_\text{QH}=\frac{6}{13}$ could equally well indicate the non-Abelian anti-Pfaffian ($\ell=-3$) or Abelian anti-331 ($\ell=-4$) pairings. Similarly, the $\nu_\text{QP} = \frac{7}{13}$ daughter would indicate either the non-Abelian Moore-Read ($\ell=1$) or Abelian 331 ($\ell=2$) pairings. In general, each daughter has one non-Abelian single-component and one Abelian multi-component parent; see Fig.~\ref{fig.daughter_states}. When only a single daughter is observed, discriminating between single and multi-component can thus lead to a unique prediction for the parent. In GaAs systems, measurements in an in-plane magnetic field are typically sufficient to obtain this information. In particular, the $\nu=\frac{5}{2}$ plateau is known to be spin-polarized,~\cite{Feiguin_spin_2009} but its nature may be affected by disorder, as we discuss Sec.~\ref{sec.disorder}. In multilayer graphene, the valley degree of freedom provides an alternative option for multi-component states, which is more challenging to assess.

\subsection{Disorder-induced Daughters}\label{sec.disorder}
We briefly turn to the case of $\nu=\frac{5}{2}$ in narrow GaAs quantum wells. It is the oldest and best-studied even-denominator state and is observed to be either childless or with a single, relatively feint descendant. At this filling, numerical works indicate that energetics favor Moore-Read and anti-Pfaffian nearly equally, with Landau-level mixing serving as a tie-breaker.~\cite{Morf_transition_1998,Rezayi_incompressible_2000,Peterson_Finite_Layer_Thickness_2008,Wojs_landau_level_2010,Storni_fractional_2010,Rezayi_breaking_2011,feiguin_density_2008,Feiguin_spin_2009} The thermal transport signatures could then arise from a PH-Pfaffian that forms at longer length scales due to disorder.~\cite{Mross_theory_2018,Wang_topological_2018,Lian_theory_2018} Under these conditions, would the daughters be determined by the global phase or by local energetics?

We expect that the energetics that favor Moore-Read and anti-Pfaffian would also favor their daughters. One could then envision a disorder-driven scenario similar to the one for the parent states. Local puddles of Moore-Read or anti-Pfaffian daughters could give rise to a macroscopic PH-Pfaffian daughter. Within the composite-fermion framework, disordered parents and daughters can be analyzed on equal footing. In both cases, the interface hosts two complex co-propagating composite fermion edge modes. At a transition between Moore-Read and anti-Pfaffian, or between their daughters, these modes percolate, and localization effects can realize intermediate plateaus.

However, there is one crucial distinction: Unlike their parents, the daughters arise at different filling factors and do not compete directly. To form puddles, the disorder would have to generate a local density imbalance to vary the filling factor on the order of 1$\%$. In the ultra-clean samples where the $\nu=\frac{5}{2}$ plateau is observed, we consider this possibility unlikely. In principle, we would, therefore, expect \text{both} sets of daughters on either side of a disorder-driven PH-Pfaffian plateau. Observing these states would provide substantial support for the disorder-induced PH-Pfaffian scenario. Still, interactions that favor both states similarly will not be ideal for either, which could explain the weakness or absence of daughters around $\nu=\frac{5}{2}$ in GaAs.

\subsection{Daughters at quarter-filling}
Daughter states are not an exclusive feature of half-filled quantum Hall states. As we discussed in Sec.~\ref{sec.next_gen}, they can occur flanking any even-denominator state whose gap is caused by pairing of composite fermions. The strongest even-denominator plateaus apart from half-filling arise at $\nu=\frac{1}{4}$ in wide quantum wells,~\cite{Luhman_1_4_wide_QW_2008,Shabani_asym_wide_QW_2009,Shabani_wide_QW_2009} and $\nu=\frac{3}{4}$ in hole-doped GaAs systems.~\cite{Wang_even_3_4_2022,Wang_Next_generation_2023}

If the pairing channel of a $\nu=\frac{1}{4}$ is Moore-Read ($\ell=1$), the daughter-state filling factors are $\nu_\text{QH}=\frac{8}{33}$ and $\nu_\text{QP}=\frac{7}{27}$; those of anti-Pfaffian ($\ell=-3$) are $\nu_\text{QH} =\frac{6}{25}$ and $\nu_\text{QP}=\frac{9}{35}$. As in the half-filled case, observing anomalously strong plateaus compared to the Jain states of four-flux composite fermions at $\nu=\frac{n}{4n\pm 1}$ would indicate the pairing channel. 

Plateaus at $\nu=\frac{3}{4}$ can equivalently be described as particle-hole conjugates of the $\nu=\frac{1}{4}$ state or as superfluids of second-generation composite fermions in Sec.~\ref{sec.next_gen}. We use the latter to assign a pairing channel $\ell$ to the resulting topological orders. Within this convention, the daughters of the Moore-Read pairing ($\ell=1$) correspond to fillings $\nu_\text{QH}=\frac{25}{33}$ and $\nu_\text{QP}=\frac{20}{27}$; the anti-Pfaffian ($\ell=-3$) parent would conceive daughters at $\nu_\text{QH}=\frac{19}{25}$ and $\nu_\text{QP}=\frac{26}{35}$. Notice that for $\nu=\frac{1}{4}$ and $\nu=\frac{3}{4}$, particle-hole symmetry relates orders with identical pairing channels. In particular, the $\nu=\frac{3}{4}$ anti-Pfaffian edge comprises three \textit{downstream} Majorana modes and a pair of counter-propagating bosons for a thermal Hall conductance of $\kappa_{xy}=\frac{3}{2}\kappa_0$.

\section{Conclusions}\label{sec.conc}

We have derived and characterized the daughter states of even-denominator quantum Hall plateaus with arbitrary pairing. Our analysis has moreover revealed their origin in strong pairing between composite fermions. The biggest question regarding these states is whether their presence constitutes reliable evidence for the nature of their parent state. Based on the picture developed here, we expect that this is indeed the case. However, numerics and experiments would be needed to put this hypothesis on firmer ground. Our findings provide guidance for both.

Experimentally, one needs to independently identify the pairing channel of the even-denominator states and compare it to the filling factor of satellite plateaus. Meanwhile, any putative daughter state needs to be distinguished from the Jain state at the same filling factor to confirm her origin in pairing. 

Various experimental signatures of different pairing channels of parent states, such as thermal conductance, upstream noise, and charge transport in interface geometries, have been discussed in Refs.~\onlinecite{Spanslatt_Noise_2019,Park_Noise_2020,Yutushui_Identifying_2022,Manna_Full_Classification_2022,Spanslatt_Noise_2022,Yutushui_Identifying_non_abelian_2023,Manna_Experimentally_2023}. In GaAs, measurements of the thermal conductance~\cite{Banerjee_Observed_2017,Banerjee_observation_2018,Dutta_Isolated_2022,Melcer_Heat_Conductance_2023} and upstream noise~\cite{Bid_Observation_neutral_2010,Dutta_novel_2022} point toward PH-Pfaffian pairing. In other platforms, there have not been any experiments that measure the pairing channel directy. 

In contrast to parent states, the daughters do not compete among themselves as they occur at different filling factors. Thus, the identification of any daughter state requires ruling out only one Jain state. This is an easier prospect since the two contenders differ substantially. In particular, their thermal Hall conductances exhibit a large difference $|\kappa^\text{Daughter}_{xy} - \kappa^\text{Jain}_{xy}|=8\kappa_0$, compared to $\frac{\kappa_0}{2}$ required to distinguish neighboring even-denominator states with $\ell$ and $\ell+1$. Moreover, all experimentally observed daughters can be distinguished from aliased Jain states with upstream noise on the vacuum interface. For the unique identification of the parent, two interfaces are needed.\cite{Yutushui_Identifying_non_abelian_2023} Finally, some experimentally observed daughter states at $\nu_\text{QH} = \frac{8}{17}$ and $\nu_\text{QP} = \frac{9}{17}$ could be identified via shot noise of tunneling through vacuum. The single-electron states are gapped on the edge when it is reduced to a topological minimum. Thus, the charge that can tunnel into the edge is $e^*=2$ and $e^*=3$, respectively.

On the numerical side, studies of daughter states are scarce. A notable exception is the filling $\nu=\frac{6}{13}$, for which a parton wavefunction was proposed and analyzed in Ref.~\onlinecite{Balram_parton_6_13_2018}. The $K$-matrix of this state [Eq.~(5) in Ref.~\onlinecite{Balram_parton_6_13_2018}] is related by a basis transformation to our Eq.~\eqref{eq.K_QH} with $m=-2$ for QH. Additionally, a recent numerical study at $\nu=\frac{8}{17}$ explored a possible competing phase to both Jain and daughter states at the same filling factors.~\cite{balram_nodaughter_2024} However, a direct comparison is impeded by the unavailability of trial wavefunctions for the latter.

 The large numerators of the daughter state filling factors severely restrict the applicability of exact diagonalization methods. In particular, extrapolation to the thermodynamic limit is challenging due to the limited accessible system sizes. Additionally, the fractional shifts ${\cal S}$ of many daughter states prevent simple `parton' wavefunctions that can be efficiently studied using Monte Carlo techniques.~\cite{Jain_Incompressible_1989,Balram_parton_2018,Balram_parton_aRR_2019,Yutushui_Large_scale_2020,Balram_parton_4_11_2021,Sharma_unconventional_2023} Among the original Levin-Halperin states, a wavefunction amenable for numerical studies is only known for $\nu=\frac{6}{13}$. Our analysis has found integer shifts for several additional daughter states, which allows such states to be simulated efficiently. In particular, one daughter of the SU(2)$_2$ topological order, which arises at $\nu_\text{QP}=\frac{6}{11}$, has shift ${\cal S}=6$. The SU(2)$_2$ corresponds to an $f$-wave pairing of composite fermions and could be relevant for the $n=3$ Landau level of monolayer graphene or wide quantum wells at quarter filling.~\cite{Kim_Even_Denominator_f_wave_2019,Faugno_Prediction_2019,Sharma_unconventional_2023} For this state, we propose the parton ansatz $(321)$, i.e., the wavefunction
\begin{align}
  \Psi^\text{QP}_{\frac{6}{11}} = \phi_3\phi_2\phi_1~,\label{eqn.123}
\end{align}
which correctly produces the filling factor, the shift, and thermal conductance $\kappa_{xy}=4$. This wavefunction can be efficiently sampled with Monte-Carlo techniques, and its real-space entanglement spectra~\cite{Sterdyniak_RES_2012,Dubail_RES_2012} are accessible with the entanglement wavefunction technique.~\cite{Rodriguez_RES_2013,Anand_RES_2022,Anand_RES_2023} 

Further investigation of Eq.~\eqref{eqn.123} and other wavefunctions of daughter states could confirm their topological order and provide insights into which interactions stabilize these phases. The physical picture of composite fermion pairs forming IQH states may also lead to alternative trial wavefunctions beyond the parton paradigm. The ingredients for forming daughter states are composite-fermion pairing and IQH states of bosons and fermions -- all of which are well understood individually. 

Finally, we briefly want to look beyond even denominators. There are two well-known classes of non-Abelian states. Firstly, there is a hierarchy of states introduced by Bonderson and Slingerland.~\cite{Bonderson_hierarchy_2008} These states exhibit pairing similar to the even-denominator states analyzed here. We thus anticipate that its daughter states also arise when those pairs form bosonic quantum Hall states. Secondly, the Read-Rezayi sequence generalizes pairing to the formation of clusters of $k\geq 2$ particles.~\cite{Read_Beyond_1999} It appears plausible that its descendants are again IQH states of $k$-particle clusters.

\textit{Note added:} In parallel to our work, Ref.~\onlinecite{Zheltonozhskii_daughters_2024} also studied daughter states and agrees with our analysis.
\begin{acknowledgments}
It is a pleasure to acknowledge illuminating conversations with Ajit Balram and Yuval Ronen on the nature of the implications of daughter states. This work was supported by the Israel Science Foundation (ISF) under grant 2572/21 and by the
Minerva Foundation with funding from the Federal German
Ministry for Education and Research. MH acknowledges funding by the Knut and Alice Wallenberg Foundation under grant no. 2017.0157
\end{acknowledgments}
\appendix

\section{Upstream noise signature}\label{app.noise}

In the main text, we asserted that some of the first-generation daughters could be distinguished from the Jain states at the same filling factor by upstream noise on the vacuum interface. Such noise implies the presence of upstream modes.~\cite{Bid_Observation_neutral_2010,Dutta_novel_2022} In this appendix, we carefully analyze the chirality of the daughter states and compare it to the Jain states. Furthermore, we show that measuring upstream noise on the interfaces with experimentally accessible reference Jain states would be sufficient to identify {\it all} first-level daughters. 

\subsection{Chirality of the QH daughter states}
 
The QH daughter states at $\nu_\text{QH} = \frac{8+m}{2(8+m)+1}$ are described by the $m+2$-dimensional $K$-matrix 
\begin{align}\label{eq.app_KQH}
  K_\text{QH} = 2\,\vect{t}\vect{t}^T + K_+,\quad \vect{t} = (2,2,1,\ldots,1)~,
\end{align}
with $K_+ = \text{diag}(\sigma_x,s,\ldots,s)$ and with $s=\text{sgn}(m)$. Since $K_+$ is a particle-like state ($\nu_+>0$ for $m>-8$), the flux attachment does not change its chirality. The boson IQH part $\sigma_x$ contributes two counter-propagating modes, while the remaining fermionic IQH supplies $|m|$ chiral 
 upstream (downstream) modes for $m>0$ ($m<0$). Some of these modes may be topologically unprotected and subject to localization. If there are protected upstream modes, noise measurements provide a sharp distinction from Jain states, which are chiral at the filling factor $\nu_\text{QH}$.

As we showed in Sec.~\ref{sec.trion}, if the $K$-matrix has a block $K_0=\text{diag}(\sigma_x,1)$ with $\vect{t}_0=(2,2,1)$, it is topologically unstable and reduces to $K^0_\text{red}=1$ with $\vect{t}_\text{red}=(3)$. For $\nu_\text{QH}$ states, this situation occurs for $m\geq 1$. The numbers $(n_\text{up},n_\text{down})$ of upstream and downstream modes are modified as $(m+1,1)\to(m-1,0)$, resulting in a chiral edge. In this case, the daughter states are indistinguishable from Jain states. By contrast, all states with $m<1$ exhibit upstream modes. In particular, both experimentally observed QH daughter states, for which $m=-2$ and $m=0$, can be distinguished from Jain states on the interfaces with vacuum.

\subsection{Chirality of the QP daughter states}
The QP daughter states at $\nu_\text{QP} = \frac{8-m}{2(8-m)-1}$ are similarly described by
\begin{align}
  K_\text{QP} = 2\,\vect{t}\vect{t}^T + K_-~,
\end{align}
with $K_-= \text{diag}(-\sigma_x,s,\ldots,s)$ and the same $\vect{t}$ as in the $\nu_\text{QP}$ case. However, in this case, $\nu^* = \vect{t}^TK^{-1}_-\vect{t}<0$ for $m<8$, and flux attachment alters the chirality of the edge. The composite-fermion $K$-matrix $K_-$ has $(m+1,1)$ for $m\geq0$ and $(1,|m|+1)$ for $m<0$ upstream and downstream modes. By contrast, the electron $K$-matrix $K_\text{QP}$ has $(m+2,0)$ and $(2,|m|)$, respectively. All states with $m\geq 0$ can thus be distinguished from Jain states, which are non-chiral at filling factor $\nu_\text{QP}$.

For $m<0$, there is a block $K_0=-\text{diag}(\sigma_x ,1)$ that admits localization in the presence of disorder. The numbers of modes reduce according to $(2,|m|)\to(1,|m|-1)$. Consequently, the QP daughter with $m=-1$, which arises at $\nu_\text{Jain}=\frac{9}{17}$, can also be distinguished from the corresponding Jain state. Importantly, both observed QP daughter states, for which $m=\pm1$ can be identified with upstream noise on the vacuum interface.

\subsection{Universal identification of daughter with upstream noise}
As we have seen, not all daughter states can be distinguished from Jain states at the vacuum interface. In those cases, interfaces with different Jain states allow such differentiation. The interfaces between any two Jain states at $\nu_n=\frac{n}{2n\pm 1}$ and $\nu_{k}=\frac{k}{2k\pm1}$ are chiral with $|n-k|$ downstream modes. Consequently, a daughter state is indicated by upstream noise at the interface with any Jain state on the same side of $\nu=\frac{1}{2}$.

\paragraph{The QH daughters}
The interface of QH daughter state at $\nu_\text{QH} = \frac{8+m}{2(8+m)+1}$ with the Jain state at $\nu_{k}=\frac{k}{2k+1}$ is described by
\begin{align} 
  K_\text{QH-$k$} = \begin{pmatrix}
  p\,\vect{t}\vect{t}^T + K_+&0\\0 & -\left[p\,\vect{t}_k\vect{t}_k^T + K^{k}_\text{IQH}\right]
  \end{pmatrix}~,
\end{align}
with $p=2$ and $\text{dim}(K^{k}_\text{IQH})= k$ and $\vect{t}_k=(1,\ldots,1)$ representing Jain states. The topological stability can be deduced from composite fermion edge with $p=0$. For $m\geq 0$, the $m$ fermionic integer modes of $K_+$ and $k$ integer modes of $K^{k}_\text{IQH}$ counter-propagate and maximally localize each other. 

If we choose $k\geq m$, the edge has $(1,1+k-m)$ modes after localization. Moreover, for $k<8+m$, the filling factor difference across the interfaces is positive, $(\vect{t},\vect{t}_k)^TK_\text{QH-$k$}(\vect{t},\vect{t}_k)>0$, and flux attachment does not change the chirality of the modes. Consequently, the interface with $k \in[m,m+7]$ is non-chiral and permits a sharp distinction between daughter and Jain states. For example, the QH daughter state at $\nu_\text{QH} = \frac{9}{19}$ can be identified using its interface with the $\nu=\frac{1}{3}$ ($k=1$) Jain states.

\paragraph{The QP daughters}
The interface of a QP daughter state and the Jain state at $\nu_{k}=\frac{k}{2k-1}$ is described by
\begin{align} 
  K_\text{QP-$k$} = \begin{pmatrix}
  p\,\vect{t}\vect{t}^T + K_-&0\\0 & -\left[p\,\vect{t}_k\vect{t}_k^T - K^{k}_\text{IQH}\right]
  \end{pmatrix}~.
\end{align}
Following the same steps as for QH, we find a non-chiral interface between $m<0$ daughter states with Jain states at $k\in[|m|,|m|+7]$. For example, to distinguish the QP daughter state at $\nu_\text{QP} = \frac{10}{19}$, one can take interfaces with $\nu=\frac{2}{3}$ ($k=2$).

\section{$K$-matrix for an Abelian superfluid with arbitrary pairing}\label{app.K_SF}
In the main text, we constructed the $K$-matrix for an Abelian superfluid of composite fermions with arbitrary pairing $\ell=2m$ by attaching $p$ fluxes to $\tilde{K}^p_\ell = p\,\vect t \vect t^T +K^*$ with
\begin{align}
    K^* = \text{diag}(K_\text{SF},K^{m-1}_{IQH}), \quad \vect t^T=(1,\ldots,1),
\end{align}
where $K_\text{SF} = \mathbb{I}-\sigma_x$, and $|m-1|$-dimensional $K^{m-1}_{IQH} = \text{sgn(m-1)}\text{diag}(1,\ldots,1)$. This $K$-matrix correctly describes quasiparticle content in accordance with, e.g., Ref.~\onlinecite{Ken_sixteenfold_2019}. However, for negative $m-1$, the $K_\ell$ is not topologically stable and admits reduction Ref.~\onlinecite{Haldane_Stability_1995}. After the reduction, the topological order is represented by the T-stable $|m|+1$ dimensional $K$-matrix Eq.~\eqref{eq.K_ell} for any non-zero $m$. Here, we show that in the $m=0$ case, Eq.~\eqref{eq.K_ell_0} results in the $K=8$ topological order after the reduction of Eq.~\eqref{eq.K_ell}. 

\subsection{The $K=8$ example}
For the $m=0$ case, the $K$-matrix given by Eq.~\eqref{eq.K_QH} with Eq.~\eqref{eq.K_ell_0} is  
\begin{align}
    K_0 = \begin{pmatrix}
    3 & 1 & 2\\
    1 & 3 & 2\\
    2 & 2 & 1
 \end{pmatrix}~.
\end{align}
The null-vector $\vect{l}=(1,-1,1)$ indicates that $K_0$ represents a topologically unstable phase. This becomes evident by reducing $K_0$ to primitive form  $K_0'=W^TK_0W$ and $\vect{t}_0'~=~W^T\vect{t}=(2,1,0)$ by $W\in\text{SL}(3,\mathbb{Z})$ with
\begin{align}
    W= \begin{pmatrix}
    1 & 0 & 0\\
    -3 & -1 & 1\\
    4 & 2 & -1
 \end{pmatrix},\quad K_0'=\begin{pmatrix}
    8 & 0 & 0\\
    0 & -1 & -1\\
    0 & -1 & 0
 \end{pmatrix}.
\end{align}
After the topological reduction, the last two fields decouple, and the resulting topological order is given by $K=8$ with $t=2$.~\cite{Haldane_Stability_1995} Thus, we immediately recognize the familiar $K=8$ topological order.

\section{Daughter states of $\nu=\frac{1}{2p}$ from CFT}\label{app.non-abelian}
In the main text, we derived the daughter-state filling factor for half-filled plateaus from the non-Abelian quasiparticle condensation. The filling factor $\nu=\frac{1}{2p}$ is related to the $\nu=\frac{1}{2}$ case by a different compactification radius of the chiral boson $\varphi$ responsible for the charge. The electron and quasihole operators are 
\begin{align}
 \psi_{e,0}(z) &= \gamma e^{i\sqrt{2p}\varphi}~,\quad
 \psi_{e,\pm k}(z) = e^{i\sqrt{2p}\varphi \pm i\phi_k}~,
 \\
 \chi_{\vect{s}}(w)& = \sigma e^{i\varphi/\sqrt{8p}} \prod^{m}_{k=1}e^{i s_k \phi_k/2}~.
 \end{align} 

An electron wavefunction can, in principle, be written as a CFT correlator of the combination of the electron operators $\psi_{e,k}$ and their descendant fields. The precise combinations that give energetically reasonable wavefunctions are known only for two cases, $m=0$ and $m=\pm 1$. For $m=0$, there is a single electron operator given by $\psi_\text{e}=\gamma e^{i\sqrt{2p}\varphi}$. For $m=1$, the $j=1$ triplet $\psi_{j_z}$ of SU(2)$_2$ should form the combination $\psi_\text{e} = (\psi_{-1} +2\sqrt{2}\partial_z \psi_0 +4\partial_z^2 \psi_{1})e^{i\sqrt{2p}\varphi}$ to give a viable electron wavefunction.~\cite{Henderson_Conformal_2023}

\subsection{Quasihole daughter state}
Suppose we know the combination of electron operators $\psi_e$. Then, the wavefunction of electrons in the presence of quasihole at $w_a$ can be written as
\begin{align}\label{eq.wf_el_app}
 \Psi_\alpha^{\{\vect{s}_a\}}(z_i;w_a) = \left\langle
 \prod_{i}\psi_{e}(z_i)
 \prod_{a}\chi_{\vect{s}_a}(w_a)
 \right\rangle_\alpha~.
\end{align}
As before, we assume that the quasiparticles of the same type, i.e., $\vect{s}_a=\vect{s}$ for all $a$, are condensed, and we omit the label $\vect{s}$. The complex conjugate of the quasihole wavefunction $\bar{\Phi}^\text{QH}_\alpha(\bar{w}_a)$ should have the opposite transformation properties under particle exchange as Eq.~\eqref{eq.wf_el_app}. This ensures that the wavefunction of the daughter state 
\begin{align} 
\Psi^\text{QH}_\text{Daughter}(z_i)=
 \sum_{\alpha}\int d w\;
 \bar{\Phi}^\text{QH}_\alpha(\bar{w}_a)
 \Psi_\alpha(z_i;w_a)
\end{align}
does not have branch cuts and is a valid electron wavefunction. Such a quasihole wavefunction can be written as a CFT correlator of operators 
\begin{align}\label{eq.qh_op_app}
\xi_{\vect{s}}(\bar{w}) = \bar{\sigma}(w) e^{i\,\sqrt{2n+\frac{1}{8p}}\bar{\varphi}(\bar{w})}\prod^{m}_{k=1}e^{i s_k \bar{\phi}_k(\bar{w})/2}~.
\end{align}
The quasihole wavefunction is then given by
\begin{equation}
\begin{split}
 \bar{\Phi}^\text{QH}_\alpha(\bar{w_a})& = \left\langle
 \prod_{a}\bar{\xi}_{\vect{s}}(\bar{w}_a)
 \right\rangle_\alpha \\
 &= 
 \langle\bar{\sigma}(\bar{w}_1)\ldots\rangle_\alpha \prod_{a<b}\bar{w}_{ab}^{2n+\frac{1}{8p} + \frac{m}{4}}.
 \end{split}
\end{equation}
Here, we used that all quasiholes are of the same type and, consequently, $\sum_{k=1}^{m}s_k^as^b_k=\sum_{k=1}^{m}s_k^2=m$ for all $a$ and $b$. We immediately infer the anyonic filling fraction $\nu_\text{QH}^\text{anyon}$ 
\begin{align}\label{eq.nu_qha_app}
 \nu^\text{anyon}_\text{QH} = \frac{1}{2n+\frac{1}{8p} + \frac{\ell-1}{8}} \qquad \text{(Anyon filling)}.
\end{align}
The electron filling factor is obtained by noting that charge density is reduced $\rho = \rho_0 - q\rho_\text{QH}$ due to the presence of quasiholes, where $\rho_0=\frac{1}{2p}B$ is the unperturbed electron density, $q=\frac{1}{4p}$ is charge of quasihole in units of electron charge, and $\rho_\text{QH} = \nu^\text{anyon}_\text{QH} B^*$ is the number density of quasiholes. Since quasihole feels an effective magnetic field $B^*=qB$, the electron filling factor $\nu_\text{QH}=\frac{\rho}{B}$ of daughter states obtained by quasihole condensation is 
\begin{align}\label{eq.nu_qh_app}
 \nu_\text{QH} =\frac{1}{2p} - \frac{1}{(4p)^2}\nu^\text{anyon}_\text{QH} \qquad \text{(Electron filling)}.
\end{align} 

\subsection{Quasiparticle daughter state}
Similar to the half-filled case, we take the wavefunction of quasiparticles to be the CFT correlator of 
\begin{align}
\rho_{\vect{s}}(w) = \sigma'(w) e^{i\,\sqrt{2n-\frac{1}{8p}-\frac{1}{4}}\varphi'(w)} \prod_{k=1}^{m} e^{is_k\bar{\phi}_k(\bar{w})/2}~.
\end{align}
To reverse the monodromy, we closely follow Levin and Halperin Ref.~\onlinecite{Levin_daughter_2009}. We similarly offset the exponent of $\varphi'$ and define the same $R_{\beta\alpha}$ as in Eq.~(20) of Ref.~\onlinecite{Levin_daughter_2009}. At the same time, the branch cuts arising from the last term of Eq.~\eqref{eq.chi} are not altered if $\vect{s}\to-\vect{s}$ and do not require modification. Thus, the valid quasiparticle wavefunction 
\begin{align}
 \Phi^\text{QP}_\alpha(w_a) = P_\text{LLL} \langle\sigma'(w_1)\ldots\rangle_\beta R_{\beta\alpha}\nonumber\\ \times \prod_{a<b} w_{ab}^{2n-\frac{1}{8p}-\frac{1}{4}} \bar{w}^{\frac{m}{4}}_{ab}~,
\end{align} 
when projected to the lowest Landau level (the factors $\prod_{a<b}|w_{ab}|^x$ can be ignored) corresponds to the anyonic filling factor
\begin{align}\label{eq.nu_qpa_app}
\nu^\text{anyon}_\text{QP} = \frac{1}{2n-\frac{1}{8p}-\frac{\ell+1}{8}} \qquad \text{(Anyon filling)}.
\end{align}
Thus, the daughter state obtained by condensation of quasiparticles occurs at the electron filling factor
\begin{align}\label{eq.nu_qp_app}
 \nu_\text{QP} = \frac{1}{2p} + \frac{1}{(4p)^2}\nu^\text{anyon}_\text{QP} \qquad \text{(Electron filling)}.
\end{align}

\section{Wire models in Abelian bosonization}\label{app.wire}
Here, we treat the wire model from the main text with Abelian bosonization.~\cite{Kane_wires_2002,Teo_LL_to_non_Abelian_2014} The non-interacting Hamiltonian density Eq.~\eqref{eq.H_nonint} on the $y$th wire takes the form
\begin{align}
 H_\text{kin} = \frac{v_F}{2\pi}\sum_{\sigma}[(\partial_x\varphi_{\sigma,y})^2+(\partial_x\theta_{\sigma,y})^2]~,
 \label{eqn.app.h0}
\end{align}
where the bosonic fields obey the usual commutation relation $[\partial_x\theta_{\sigma,y}(x),\varphi_{\sigma,y}(x')] = i\pi \delta(x-x')$. The electron operators are expressed as
\begin{align}
 \psi^R_{\sigma,y}&\sim e^{i(\phi^R_{\sigma,y} + k^\sigma_{F}x)}= e^{i(\varphi_{\sigma,y}+\theta_{\sigma,y} + k^{\sigma}_{F}x)}~,\\ 
 \psi^L_{\sigma,y}&\sim e^{i(\phi^L_{\sigma,y} - k^\sigma_{F}x)}= e^{i(\varphi_{\sigma,y}-\theta_{\sigma,y}  -k^{\sigma}_{F}x)}~,
\end{align}
where $k^\sigma_F = \rho_\sigma \pi a$ is determined by the density $\rho_\sigma$ of $\sigma$ electrons, with $a$ the distance between wires. The magnetic field enters the tunneling terms through $b=|e|aB/\hbar$. The filling factor then reads $\nu=2(k^\up_F+k^\down_F)/b$.

\subsection{Superfluids and quantum Hall states with $m=0$}
 In the boson variables, the Cooper-pair hopping Hamiltonian of Eq.~\eqref{eqn.hpair} and Josephson coupling Hamiltonians of Eq.~\eqref{eqn.hm} are given by 
\begin{align}
 H_\text{pair} &= g\cos[\phi^R_{\up,y+1} + \phi^L_{\down,y+1}-\phi^R_{\up,y} -\phi^L_{\down,y}]~,\\
 H_{m=0} &= g_0\cos[\phi^L_{\up,y} + \phi^R_{\down,y}-\phi^R_{\up,y} -\phi^L_{\down,y}]~,
\end{align}
where we adjusted the density to eliminate phase factors. To analyze these terms, we introduce charge and spin modes according to
\begin{equation}
\begin{split}
  \varphi_c &= \varphi_\up+\varphi_\down
   ~,\quad 2\theta_c =\theta_\up+\theta_\down~,\\
   \varphi_s &= \varphi_\up-\varphi_\down
     ~,\quad 2\theta_s =\theta_\up-\theta_\down
     ~.
 \end{split}
 \end{equation}
These modes satisfy the canonical commutation relations $[\partial_x\theta_{q,y}(x),\varphi_{q',y}(x')] = i\pi\delta_{qq'}\delta(x-x')$ with $q=c,s$.

The Josephson coupling term takes the form
\begin{align}
 H_{m=0} &= g_0\cos[4\theta_{s,y}]~,
\end{align}
and pins each $\theta_{s,y}$ to one of its minima when relevant. Upon dropping these pinned fields, the Cooper-pair hopping terms becomes
\begin{align}
   H_\text{pair} &= g\cos[\varphi_{c,y+1}-\varphi_{c,y}]~.
\end{align}
In the strong coupling limit where $g$ is of order unity, this term still permits smooth variations of $\varphi_c$ at low energies. These fluctuations describe the Goldstone mode of the superfluid. A detailed discussion of such modes in coupled wire systems can be found in Ref.~\onlinecite{leviatan_wires_2020}.

\paragraph{Flux attachment}
To implement flux attachment according to Refs.~\onlinecite{Mross_explicit_duality_2016,Mrossduality2017,fuji_wires_2019,leviatan_wires_2020}, we infinitesimally shift the $\down$ wires to $y \rightarrow y+ 0^+$ and the $\up$ wires to $y \rightarrow y+ 0^-$. The attachment of $p$ flux quanta is then achieved by the replacement
\begin{equation}
\begin{split}
 \varphi_{\up,y} =\varphi_{\up,y} + 2p\sum_{y'\neq y} \text{sgn}(y-y')\theta_{c,y} + 2p\theta_{\down,y}~,\\
\varphi_{\down,y} =\varphi_{\down,y} + 2p\sum_{y'\neq y} \text{sgn}(y-y')\theta_{c,y} - 2p\theta_{\up,y}~,
\end{split}
 \end{equation}
without modifying $\theta_\uparrow$ or $\theta_\downarrow$. In the charge-spin basis, the flux-attachment transformation reads
\begin{equation}
\begin{split}
  \varphi_{c,y} &\to \varphi_{c,y} + 4p\sum_{y'\neq y} \text{sgn}(y-y')\theta_{c,y} - 4p\theta_{s,y}~,\\  
  \varphi_{s,y} &\to \varphi_{s,y} + 4p\theta_{c,y}~,
\end{split}
 \end{equation}
and $\theta_{c/s}$ are unaffected. 

The Josephson coupling term for $m=0$ is unchanged, and $\theta_{s,y}$ are again pinned. Dropping these constants, we find that the Cooper-pair hopping is replaced by
\begin{align}
   \tilde H_\text{pair} &= g\cos[\varphi_{c,y+1}-\varphi_{c,y} -4p(\theta_{c,y+1} + \theta_{c,y})]~.
\end{align}
The Hamiltonian $\tilde H_\text{pair}$ was shown in Refs.~\onlinecite{Kane_wires_2002} to realize a $\tilde\nu_\text{Pair}=\frac{1}{4p}$ Laughlin state of the bosons $e^{i \varphi_c}$. Since this boson carries charge $e_\text{Pair}=2$, the electron filling factor of this state is $\nu=\frac{1}{p}$. Attaching $p=2$ flux quanta thus yields the $K=8$ state of electrons.

\subsection{Superfluids with $m\neq 0$}
For general $m$ it is convenient to adopt the $m$-dependent basis
\begin{align}\label{eq.app_varphi}
 2\varphi^1_{y} &= \phi^R_{\up,y} + \phi^L_{\down,y},
 \qquad
 2\theta^1_{y} = \phi^R_{\up,y} - \phi^L_{\down,y}~,\\
   2\varphi^2_{y}&=\phi^L_{\up,y} + \phi^R_{\down,y- m},\quad
  2\theta^2_{y}=\phi^L_{\up,y} - \phi^R_{\down,y- m}~.
\end{align}
The Cooper-pair hopping of Eq.~\eqref{eqn.hpair} and Josephson coupling of Eq.~\eqref{eqn.hm} take the form
\begin{align}
 H_\text{pair} &= g\sum_y\cos[\varphi^1_{y+1} - \varphi^1_{y}]~,\\
 H_{m=0} &= g_0\sum_y\cos[\varphi^2_{y+m} - \varphi^1_{y}]~.
\end{align}
The Hamiltonian $H_\text{pair}$ encodes a Goldstone mode as for $m=0$. The cosine in $H_{m=0}$ slaves $\varphi^2$ to $\varphi^1$ in the bulk. However, near a boundary, $m$ of the $\varphi^2$ field are masterless. As a result, a finite system of wires with $y_0\leq y \leq y_1$ exhibits $m$ chiral edge modes $\phi^L_{\up,y < y_0+m}$ near the bottom boundary and $\phi^R_{\down,y > y_1-m}$ near the top boundary.

\subsection{Quantum Hall states with $m =  1$}
Flux attachment modifies the Josephson inter-wire coupling to
\begin{align}
 H_{m=1} =
 g_0\cos[
 &\varphi_{\up,y+1}-3 \theta_{\up,y+1}
 - 2 \theta_{\down,y}
 -\varphi_{\up,y}-3 \theta_{\up,y}
]~,
 \end{align}
while the pair-hopping term becomes 
\begin{align}
 H_\text{pair} = g\cos[
 & \varphi_{\down,y+1}-3\theta_{\down,y+1}-2 \theta_{\up,y+1}-\varphi_{\down,y}-3 \theta_{\down,y}\nonumber\\
 +&\varphi_{\up,y+1} -3 \theta_{\up,y+1}-2 \theta_{\down,y}-\varphi_{\up,y}-3\theta_{\up,y}]~.
\end{align}
The lower row in $H_\text{pair}$ is pinned by the first term and can be replaced by a constant. A simply relabeling $(y,\downarrow) \rightarrow 2j+1$ and $(y,\uparrow) \rightarrow 2j$ then precisely reproduces the coupling terms in Eq.~(2.45) of Ref.~\onlinecite{Teo_LL_to_non_Abelian_2014} for the 331 state.

\subsection{Boson integer quantum Hall effect}
The boson IQH effect of Cooper pairs~\cite{Chen_SPT_2012,Lu_Classification_2012,Levin_Classification_2012,Ashvin_BIQH_2013,Senthil_BIQH_2013,Geraedts_realization_2013} is realized when $H_0$ and $H_\text{BIQH}$ flow to strong coupling. To analyze this phase, we define the charge and neutral basis
\begin{align}
 \phi^{R/L}_{c,y} = \phi^{R/L}_{\up,y} + \phi^{R/L}_{\down,y}~,\quad
 \phi^{R/L}_{n,y} = \phi^{R/L}_{\up,y} - \phi^{R/L}_{\down,y}~,
\end{align}
and corresponding non-chiral variables
\begin{align}
  2\varphi_{c/n}=\phi^{R}_{c/n} + \phi^{L}_{c/n},\qquad 4\theta_{c/n} =\phi^{R}_{c/n} - \phi^{L}_{c/n}~,
\end{align} 
which obey $[\partial_x\theta_{c/n,y}(x),\varphi_{c/n,y}]=i\pi\delta(x-x')$. We immediately see that spin degree of freedom is completely gapped by Josephson-type term
\begin{align}
 H_0 = g_0 \cos[4\theta_{n,y}]~.
\end{align}
After $\theta_{n,y}$ are pinned, the hopping of boson-vortex composites in Eq.~\eqref{eqn.bosonvortex} reduces to
\begin{align} 
 H_\text{BIQH} = \sum_y g_b \cos[\varphi_{c,y-1} - \varphi_{c,y+1} \pm 2\theta_{c,y}]~,
\end{align}
where the choice of the sign is determined by the Cooper-pair filling factor $\nu_\text{Pair}=\pm 2$. This inter-wire coupling is precisely Eq.~(1) in Supplemental Material of Ref.~\onlinecite{mross_biqh_2016}. To make a connection with the $K$-matrix, i.e., Eq.~\eqref{eq.K_daughter} in the main text, we define new fields
\begin{align}
 &\phi^{\pm}_{1,y} = \varphi_{c,2y-1},\quad &
 &\phi^{\pm}_{2,y} = \varphi_{c,2y}\pm 2\theta_{c,2y-1},&
 \\
 &\phi^{\mp}_{2,y} = \varphi_{c,2y},\quad &
 &\phi^{\mp}_{1,y} = \varphi_{c,2y-1} \mp 2\theta_{c,2y} ~.
\end{align}
Their commutator satisfies $[\partial_x\phi^{\pm}_{a,y},\phi^{\pm}_{b,y}] = 2i\pi K^{-1}_{ab} \delta(x-x')$, with $K=\pm\sigma_{x}$. 
The tunneling terms 
\begin{align}
 H_\text{BIQH} = \sum_{\alpha=1,2}\sum_y g_b \cos[\phi^{\pm}_{\alpha,y} - \phi^{\mp}_{\alpha,y\pm1}]~.
\end{align}
In a system that is finite in $y$-direction, the modes residing on the first wire $\phi^{\pm}_{\alpha,1}$ remain gapless and are described by the $K=\pm \sigma_x$ and $\vect{t}=(2,\;2)$ corresponding to the boson IQH effect at $\nu_\text{Pair}=\pm2$.

\subsection{Boson integer quantum Hall effect with $m\neq 0$}
We reduce the problem for general $m$ to the $m=0$ case by noting that $\phi^L_{\up,y}$ appears only in the Josephson-type tunneling term $H_m$. Consequently, upon defining $\tilde{\psi}^{L}_{\up,y} \equiv \psi^{L}_{\up,y+m}$, the problem can be analyzed equivalently to the $m=0$ case. The only two differences are: (i) The phase of $\psi^{L}_{\up,y+m}$ has an extra contribution $mB$; thus, the filling factor is different compared to the $m=0$ case. (ii) $m$ fields $\tilde{\psi}^{L}_{\up,y}$ with $y=-m+1,-m+2,\ldots,0$ remain gapless at the edge. 

Repeating the same analysis as for the $m=0$ case with $\tilde{\phi}^L_{\up,y}$, we find that gapless edge modes are described by $K=\text{diag}(\pm\sigma_x,1,\ldots,1)$ and $\vect{t}=(2,\;2,\;1,\ldots,1)$.

\section{Topologically reduction for $K$-matrix of BIQH}\label{app.K_BIQH}
In the main text, we constructed the $K$-matrix $K_{n_\text{Pair}}$ of boson IQH at $\nu_\text{Pair}=2n_\text{Pair}$ by appending $n_\text{Pair}$ copies of $K_1=\sigma_x$ for $\nu_\text{Pair}=2$. We now inductively show that the $K$-matrix of Eq.~\eqref{eq.K_daug_higher} reduces to Eq.~\eqref{eq.K_daug_higher_red}. 

The block-diagonal $K$-matrix $K=\text{diag}(K_{n_\text{Pair}-1},\sigma_x)$ with $\vec{t}=(2,2,2,2)$ is topologically unstable with null-vector, e.g., $\vec{l}=(0,1,-1,0)$. We transform $K$ with $W\in SL(4,\mathbb{Z})$ 
\begin{align}
    W=\left(
\begin{array}{cccc}
 1 & -2 & 1 & 1 \\
 0 & 1 & 0 & 0 \\
 0 & 1 & 0 & -1 \\
 0 & 1 & -1 & 0 \\
\end{array}
\right)
\end{align}
to $W^TKW = \text{diag}(K_{n_\text{Pair}},\sigma_x)$ and $W^T \vec{t}=(2,2,0,0)$. The $K$-matrix is now in a primitive form, and the second block cancels out, leaving behind $K_{n_\text{Pair}}$ and $\vec{t}=(2,2)$. We now inductively apply this procedure, which reduces Eq.~\eqref{eq.K_daug_higher} to Eq.~\eqref{eq.K_daug_higher_red}.

 \section{Higher-generation even denominators}\label{app.higher}
 In the main text, we derived the daughter-state filling factors for next-generation even denominator states. Such states can be described composite fermion states at $\nu^*=\pm (n+\frac{1}{2})$. The gap of the composite fermions is explained as $n$ filled Landau levels and a paired state in the half-filled $n+1$'th Landau level. The paired state, in turn, is viewed as a second generation of composite fermions forming a superconductor. In this appendix, we derive the $K$ matrices for Abelian next-generation paired states and their daughters.

\subsection{Next-generation even-denominator states}
We obtain the $K$ matrix for next-generation states at the filling factor
\begin{align}
 \nu^\text{NG}(n,p) = \frac{n+\frac{1}{2}}{p(n+\frac{1}{2}) \pm 1}~.
\end{align}
in two steps. First, we find the $K$ matrix for the first-generation composite fermions at filling $\nu^* >0$. When the composite fermions in the partially filled $n+1$'th Landau level, we take $K^*_{1/2}=K_m$ and $\vect{t}_{1/2}=\vect{t}_m$ from Eq.~\eqref{eq.K_ell}. Supplementing it by $n$ composite-fermion Landau levels yields $K^*_{\frac{2n+1}{2}}= \text{diag}(K^*_{1/2},K^{n}_\text{IQH})$ and $\vect{t}=(\vect{t}_{1/2},1,\ldots,1)$. For $\nu^* >0$, we similarly find $K^*_{-\frac{2n+1}{2}}=-K^*_{\frac{2n+1}{2}}$. To obtain the electron $K$ matrix, we attach $p$ flux quanta, i.e.,
\begin{align}\label{eq.app_flux}
  K = p\,\vect{t}\vect{t}^T + K^*_{\pm\frac{2n+1}{2}}~.
\end{align}
This $K$-matrix describes the higher-generation even-denominator states at filling factor $ \nu^\text{NG}$. The pairing channel of these higher-generation states is inherited from the $\nu^*=\frac{2n+1}{2}$ states. 

To determine the topological order of the daughter states, we proceed analogously. First, we take a daughter state of a half-filled Landau level of composite fermions given by $K_\text{QH/QP}$ from Eq.~\eqref{eq.K_QH}. Then, we extend the $K^*_\text{QH/QP}= \pm \text{diag}(K_\text{QH/QP},K^{n}_\text{IQH})$. Finally, we attach $p$ fluxes with Eq.~\eqref{eq.app_flux}.

\subsubsection{Example: Moore-Read QP daughter state of $\nu=\frac{3}{4}$} 

The QH daughter $K$-matrix is given by $K_\text{QH}$ for $m=0$ in Eq.~\eqref{eq.K_QH}. The full composite fermion $K$-matrix is 
\begin{align}
 K^* = \begin{pmatrix}
 8 & 9 & 0\\
 9 & 8 & 0 \\
 0 & 0 & 1 \\
\end{pmatrix}~,\quad \vect{t}^* = \begin{pmatrix}
 2\\
 2 \\
 1 \\
\end{pmatrix}~,
\end{align}
 which corresponds to $\nu^*_\text{QH} = 1+\frac{8}{17}$. Attaching two fluxes to a hole-like state at $-\nu^*_\text{QH}$ results in the electron state at $\nu=\frac{25}{33}$ with $K$-matrix 
\begin{align}
 K =2\,\vect{t}\vect{t}^T-K^*= \begin{pmatrix}
 0 & -1 & 4\\
 -1 & 0 & 4 \\
 4 & 4 & 1 \\
\end{pmatrix}~.
\end{align}
The eigenvalues of $K$ imply two downstream and one upstream mode, which lead to a thermal conductance of $\kappa_{xy}=1$. The $K$-matrix is topologically unstable and reduces to a single chiral mode with no upstream noise expected.

\subsubsection{Example: Anti-Pfaffian QH daughter state of $\nu=\frac{3}{4}$} 
Our second example arises from quasihole condensation on top of the $\nu=\frac{3}{4}$ anti-Pfaffian state. The same daughter state can be alternatively obtained as the QH daughter of anti-(331) topological order corresponding to $m=-2$. We obtain its $K$-matrix by two-stage flux attachment to a boson IQH state of pairs with $m=-2$ fermion IQH states, i.e., 
\begin{align}
  K^{**}_+=\left(
\begin{array}{cccc}
 0 & 1 & 0 & 0 \\
 1 & 0 & 0 & 0 \\
 0 & 0 & -1 & 0 \\
 0 & 0 & 0 & -1 \\
\end{array}
\right)~,
\end{align}
with $\vect{t}_{1/2}=(2,2,1,1)$. Flux attachment yields the first-order composite-fermion $K$-matrix 
\begin{align}
  K^*_{\text{QH},\nu^*=1/2}=\left(
\begin{array}{ccccc}
 8 & 9 & 4 & 4 \\
 9 & 8 & 4 & 4 \\
 4 & 4 & 1 & 2 \\
 4 & 4 & 2 & 1 
\end{array}
\right)~.
\end{align}
To obtain the daughter state of composite fermions at $\nu^*=-\frac{3}{2}$, we extend the $K$-matrix by one integer mode, which results in $ K^*_{\text{QH},\nu^*= -3/2} =-\text{diag}(K^*_{\text{QH},\nu^*=1/2},1)$ and $\vect{t}_{3/2}=(2,2,1,1,1)$. Finally, we attach two fluxes to obtain the electronic $K$-matrix
\begin{align}
K=\left(
\begin{array}{ccccc}
 0 & -1 & 0 & 0 & 4 \\
 -1 & 0 & 0 & 0 & 4 \\
 0 & 0 & 1 & 0 & 2 \\
 0 & 0 & 0 & 1 & 2 \\
 4 & 4 & 2 & 2 & 1 \\
\end{array}
\right)~,
\end{align} 
which corresponds to $\nu_\text{QH}=\frac{19}{25}$. The numbers of downstream and upstream modes are $(4,1)$, and the thermal Hall conductance is $\kappa_{xy}=3$, respectively. The $K$-matrix, however, is not T-stable and reduces to three downstream modes with no upstream noise.

\bibliography{phpfbib}

\end{document}